\documentclass[]{emulateapj}
\shorttitle{CO isotopologue ratios in protostellar disks and envelopes}
\shortauthors{Smith et al.}
\usepackage{multirow}

\begin{document}

\title{High-precision C$^{17}$O, C$^{18}$O and C$^{16}$O measurements in young stellar objects: analogues for CO self-shielding in the early solar system\footnote{This work is based on observations collected at the European Southern Observatory 
Very Large Telescope under program ID 179.C-0151.}}

\author{Rachel L. Smith\altaffilmark{1}, Klaus M. Pontoppidan\altaffilmark{2}, Edward D. Young\altaffilmark{1,3}, 
Mark R. Morris\altaffilmark{4}, and Ewine F. van Dishoeck\altaffilmark{5,6}}
\altaffiltext{1} {Department of Earth and Space Sciences, University of California Los Angeles, 595 Charles E. Young Drive East, 
Geology Building, Los Angeles, CA 90095-1567; rsmith@ess.ucla.edu.}
\altaffiltext{2} {Hubble Fellow, Division of Geological and Planetary Sciences, California Institute of Technology, Pasadena, 
CA 91125; pontoppi@gps.caltech.edu.}
\altaffiltext{3} {Institute of Geophysics and Planetary Physics, University of California Los Angeles; eyoung@ess.ucla.edu.}
\altaffiltext{4} {Division of Astronomy and Astrophysics, Department of Physics and Astronomy, University of California, 
Los Angeles, CA 90095-1547; morris@astro.ucla.edu.}
\altaffiltext{5} {Leiden Observatory, Leiden University, P.O. Box 9513, NL  2300 RA Leiden, The Netherlands; ewine@strw.leidenuniv.nl.}
\altaffiltext{6} {Max Planck Institut f\"{u}r extraterrestrische Physik, Postfach 1312, 85741 Garching, Germany.}

\begin{abstract}
Using very high resolution ($\lambda$/$\Delta$$\lambda$ $\approx$ 95\,000) 4.7 $\mu$m fundamental and 2.3 $\mu$m overtone ro-vibrational CO absorption 
spectra obtained with the CRIRES infrared spectrometer on the Very Large Telescope (VLT), we report detections of four CO isotopologues -- C$^{16}$O, $^{13}$CO, C$^{18}$O and the rare species, 
C$^{17}$O -- in the circumstellar environment of two young protostars, VV CrA and Reipurth 50.
We argue that the observed CO absorption lines probe a protoplanetary disk in VV CrA, and a protostellar envelope in Reipurth 50.  
All CO line profiles are spectrally resolved, permitting direct calculation of CO oxygen isotopologue ratios with 5-10\% accuracy. 
The ro-vibrational level populations for all species can be reproduced by assuming that CO absorption arises in two temperature regimes.  
For both objects, $^{12}$C/$^{13}$C are on the order of 100, nearly twice the expected interstellar medium (ISM) ratio. 
The derived oxygen abundance ratios for the VV CrA disk show a significant mass-independent deficit of C$\rm ^{17}$O and C$\rm ^{18}$O relative to C$\rm ^{16}$O compared to ISM baseline abundances. 
The Reipurth 50 envelope shows no clear differences in oxygen CO isotopologue ratios compared with the local ISM. 
A mass-independent fractionation can be interpreted as being due to selective photodissociation of CO in the disk surface due to self-shielding. 
The deficits in C$\rm ^{17}O$ and C$\rm ^{18}O$ in the VV CrA protoplanetary disk are consistent with an analogous origin of the $^{16}$O
variability in the solar system by isotope selective
photodissociation, confirmation of which may be obtained via study of additional sources. The higher fractionation observed for the VV CrA disk
compared with the Reipurth 50 envelope is likely due to a combination
of disk geometry, grain growth, and vertical mixing processes. [Abstract abridged]
\end{abstract}

\keywords{astrochemistry --- circumstellar matter --- ISM: abundances --- planetary systems: protoplanetary disks --- stars: individual (VV CrA, Reipurth 50)}

\section{Introduction}
The capability to compare isotope ratios of 
meteorites with those in protoplanetary systems presents exciting new possibilities for constraining early solar system history. Oxygen is 
of particular value in this regard because it occurs in great abundance, comprising $\sim$ 50\% by mass of most rocky materials 
and is, with the [CO]/[H] ratio roughly $10^{-4}$ in the gas phase, a significant portion of the gas as well. With the advent of improved instrumentation, astronomical observations of stable isotope ratios are becoming a potentially 
powerful new tool for probing the chemistry of young stellar objects (YSOs), making comparisons with the solar system possible.

Mass-dependent isotope fractionation describes the relationship between the changes in the ratios of three or more isotopes. In the case of oxygen, one considers the relationship between the fractionation factors $\rm \alpha^{17/16}$ and $\rm \alpha^{18/16}$, where $\rm \alpha^{17/16} = ([^{17}O]/^{16}O])/([^{17}O]/[^{16}O])_{\circ}$ and $\rm \alpha^{18/16} = ([^{18}O]/^{16}O])/([^{18}O]/[^{16}O])_{\circ}$. The subscript $\circ$ refers to some initial condition.  Because mass-dependent fractionation depends on the ratio of partition functions, it is 
well known that the relationship is $\alpha^{17/16} = (\alpha^{18/16})^{\beta}$, where $\beta = (1/m_{16}-1/m_{17})/(1/m_{16}-1/m_{18})$ at equilibrium, or 
ln$(M_{16}/M_{17})/$ln$(M_{16}/M_{18})$ for kinetic processes, and where $m_{i}$ is the atomic mass for atomic species $i$, and $M_{i}$ can 
be an atomic, molecular, or reduced mass for isotopologue $i$ depending on the process \citep{Young2002-2}.  In practice, $\beta \sim 0.51$ (kinetic)  to 0.53 (equilibrium) for oxygen.  Isotope variations that obey this 
relationship are deemed mass-dependent.  Mass-independent variations are those that do not follow this relationship.

Oxygen isotopes in rocks of the solar system exhibit an anomalous mass-independent distribution that has defied conclusive explanation since its 
discovery \citep{Clayton1973-458}. This anomaly, in the form of a mass-independent correlation 
between $\rm [^{16}O]/[^{18}O]$ and $\rm [^{16}O]/[^{17}O]$ among rocky bodies, is one of the most pronounced chemical 
features of the solar system.  Here, mass independence refers to relative differences in $\rm [^{16}O]/[^{18}O]$ and $\rm [^{16}O]/[^{17}O]$ 
among primitive rocky objects that are nearly identical, suggesting changes in $\rm [^{16}O]$ relative to both $\rm [^{17}O]$ 
and $\rm [^{18}O]$, rather than the expected mass-dependent trend in which relative changes in $\rm [^{16}O]/[^{17}O]$ are 
about half those in $\rm [^{16}O]/[^{18}O]$ on a log scale.  

Explanations for the mass-independent distribution of oxygen isotope ratios in the solar system include inheritance of a  nucleosynthesis signal from the interstellar medium (e.g., Galactic chemical evolution), symmetry-based mass-independent isotope fractionation effects such as those involving ozone in Earth's stratosphere \citep{Thiemens1983-1073}, and photochemistry involving isotope-specific photodissociation of CO \citep{Kitamura1983-192, Thiemens1983-1073, Clayton2002-860, Yurimoto2004-1763,  Lyons2005-317}.  Recent results from the Genesis mission establish that rocks in the solar system are depleted in $\rm ^{16}O$ relative to the Sun \citep{2008LPI....39.2020M}.  The depletion is extreme in aqueously altered primitive meteoritical materials (Sakamoto et al., 2007), suggesting that $\rm [H_{2}^{18}O]/[H_{2}^{16}O]$ and $\rm [H_{2}^{17}O]/[H_{2}^{16}O]$ were extremely high in the planet-forming region of the early solar system.  Both of these observations lend support to the suggestion that  self-shielding by CO oxygen isotopic species during photodissociation of CO is a likely explanation for mass-independent fractionation of oxygen isotopes in the solar system \citep{Young2008-187}. 
 
Self-shielding by CO refers to the variable shielding of CO oxygen isotopologues from photodissociation by far ultraviolet (FUV) radiation in proportion to their abundances (column densities).  Thus,  $\rm C^{17}O$ and  $\rm C^{18}O$ will be more rapidly destroyed than the much more abundant  $\rm C^{16}O$.  The oxygen liberated during this process will eventually end up in  $\rm H_{2}O$, providing an explanation for the overabundance of $\rm H_{2}^{18}O$ and $\rm H_{2}^{17}O$ relative to $\rm H_{2}^{16}O$ in the early solar system.  Exchange of oxygen isotopes between newly formed rock and $\rm H_{2}O$ gas provided a path for rocks to become enriched in $\rm^{18}O$ and $\rm^{17}O$ relative to the original solar oxygen isotopic composition \citep{Yurimoto2004-1763, Lyons2005-317, Young2007-468}. 

The effect on the survival of oxygen isotopologues is greatest for column densities of CO in the range $10^{15}$ 
to $10^{18}$ $\rm cm^{-2}$.  Isotope
selective photodissociation and self-shielding occur in molecular clouds 
 \citep{Bally1982-143,VanDishoeck1988,Sheffer2002-L171} but their
occurrence in circumstellar disks has yet to be demonstrated conclusively.  Models of CO photodissociation in a 
disk suggest that isotope selectivity will be most efficient at the surfaces of the disk 
regardless of whether the FUV source is the central star itself or proximal O or B stars 
\citep{Lyons2005-317,Young2007-468}. 
These models predict that the outer regions of circumstellar disks should 
exhibit $\rm C^{17}O$ and $\rm C^{18}O$ deficits relative to $\rm C^{16}O$ as a consequence of CO self-shielding, providing a test of the models. The precise degree of isotope selection depends on the amount of radial and vertical mixing in the disk, as well as on time. Models suggest $\rm C^{17}O$ and $\rm C^{18}O$ deficits relative to $\rm C^{16}O$ of tens of percent on time scales of $10^{4}$ to $10^{5}$ years \citep{Young2007-468}. This is more than sufficient to explain the anomalous distribution of oxygen isotopes in the solar system. An alternative model \citep{Yurimoto2002-, Lee2008-molcld} suggests that signatures of CO self-shielding in the disk may have been inherited from the surrounding cloud material, where the excess $^{17}$O and $^{18}$O has subsequently been
incorporated into water ice and transported to disks; in this case, one expects anomalous deficits in $\rm C^{17}O$ and  $\rm C^{18}O$ in the disk and in the envelope surrounding the disk, regardless of stage of evolution. 

In this  work we present column densities for the CO isotopologues $\rm ^{12}C^{16}O$ ($\rm C^{16}O$), 
$\rm ^{12}C^{18}O$ ($\rm C^{18}O$), $\rm ^{12}C^{17}O$ ($\rm C^{17}O$), and  $^{13}$C$^{16}$O ($^{13}$CO) in two YSOs 
that comprise a first test  of the importance of CO self-shielding in the early solar system.  
We emphasize that quantifying {\it both} $\rm [^{16}O]/[^{18}O]$ {\em and} $\rm [^{16}O]/[^{17}O]$ is particularly important 
in this study because only the combination of the two ratios can distinguish conclusively between photochemical 
self-shielding and mass-dependent fractionation.

\begin{figure*}
\centering
\includegraphics[width=13cm]{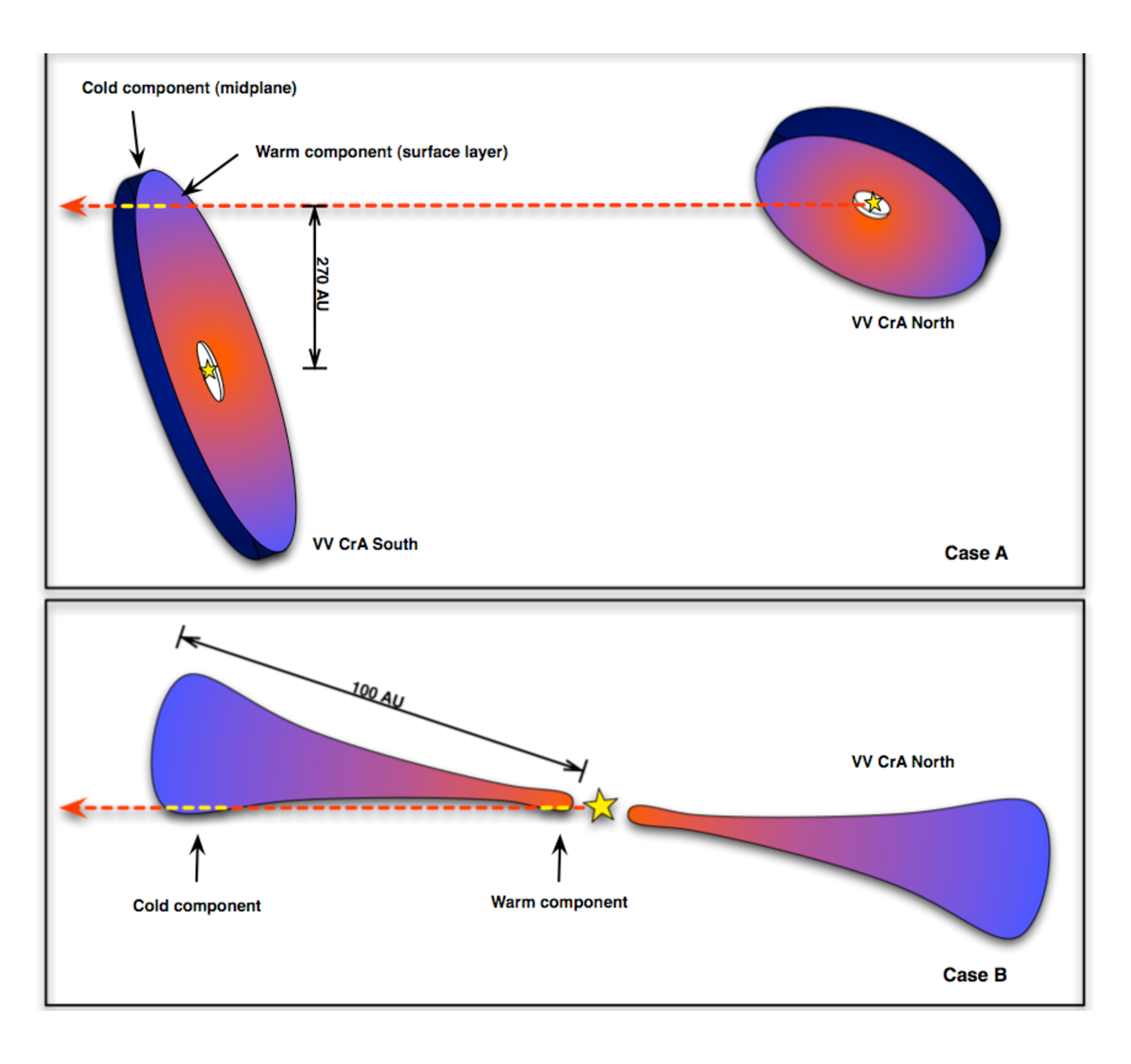}
\caption[]{Schematic cartoon illustrating the two possible geometries for the VV CrA binary system. Case A illustrates the scenario whereby the disk of the secondary star is eclipsed by the outer disk of the primary; Case B shows the less probable, single inclined disk geometry.}
\label{diskcartoon}
\end{figure*} 

Infrared absorption spectroscopy has been used previously to measure the ratios of the $^{12}$CO, $^{13}$CO and C$^{18}$O 
isotopologues in YSOs. Infrared observations of one circumstellar gaseous disk provide a hint of $\rm C^{16}O$   
overabundance relative to C$^{18}$O: \cite{Brittain2005-283} presented high-resolution infrared spectra of the embedded, low-mass 
pre-main sequence star HL Tau, showing a ratio of column densities for $\rm C^{16}O$ and 
$\rm C^{18}O$, $N({\rm C^{16}O})/N({\rm C^{18}O})$, of 800$\pm 200$ ($2\sigma$).  Typical $\rm [^{16}O]/[^{18}O]$ 
in the local interstellar medium (ISM) is $557\pm 30$, \citep{Wilson99}, similar to the solar system value of 499. 
The apparent overabundance 
of $\rm C^{16}O$ in the HL Tau disk could be the result of isotope-selective photodissociation resulting 
from the HL Tau UV field, although the uncertainties do not yet allow one to completely rule out standard ISM isotopologue ratios. Interpretation of these data are complicated by the fact that the line of sight toward HL Tau samples mostly the envelope rather than the disk. 
In addition, without the complementary data for $\rm C^{17}O$, it is not possible to differentiate between mass-independent versus mass-dependent fractionation mechanisms; at temperatures that exist in the outer regions of disks, mass fractionation could be 20\% or 
more \citep{Young2007-468}, well within the range for HL Tau.  

Searching for oxygen isotope fractionation in proto- 
planetary disks requires an observational method that 
can measure the abundances of the four most common 
CO isotopologues to relative accuracies better than 10\%. 
Isotopologue ratios measured using rotational emission 
lines depend on excitation and radiative transfer models 
with signiÞcant uncertainties due to non-thermalized level populations
and differences in beam sizes arising from differences in the frequencies
of the various CO isotopologues. Further, rotational emission is
currently limited to the lowest rotational levels $(J \leq 3)$, and is therefore
tracing low-temperature gas only. 
An alternative that eliminates most of these concerns is to use absorption lines in which 
transitions from different isotopologues are known to 
probe nearly identical pencil-beam lines of sight. The fundamental
ro-vibrational band at 4.7\,$\mu$m and the overtone at 2.3\,$\mu$m of CO provide 
such an absorption-line tracer. We show that very high resolution infrared spectroscopy
can be used to derive accurate column densities of all 3 isotopologues of $\rm^{12}CO$. 
The main challenge is to locate sources with favorable geometries providing column densities
high enough to allow the detection of C$^{17}$O, but not so high as to 
extinguish the infrared source. One type of favorable absorption-line geometry
is that of a disk viewed at high inclination 
angles to allow for absorption in the outer 
disk of infrared continuum emission originating in the 
inner disk. In this paper, we take advantage of a new type of geometry, namely
that of a binary star in which one star is being eclipsed by the disk of the other star.  
We compare one such binary T Tauri star with an embedded YSO in which the 
absorption may in part be attributed to a remnant envelope. 

\

\section{Source selection}

The first absorption-line source is the northern component of the binary T Tauri star VV CrA (J2000: RA = 19 03 06.8, dec = $-$37 12 54.6); the southern component
does not show CO in absorption. VV CrA N is part of the
class of ``infrared companion'' (IRC) sources, which are highly extincted companions to optically visible T Tauri stars with
projected separations of a few hundred AU or less \citep{Koresko97}. Infrared companions typically show silicate absorption features, revealing that the absorbing material is localized. One interpretation is that the infrared
companions have very compact, dusty envelopes. Here, we suggest a possible alternative explanation: that at least some of the infrared companions
are regular T Tauri stars that are being eclipsed by the outer disk of the primary star of the binary \citep[e.g.,][]{Hogerheijde1997}. The only strict requirement
for this scenario is that the disks of the two companions be non-coplanar, a geometry that gains support from the fact that non-coplanarity is a known property of 
some binary T Tauri stars \citep[e.g.,][]{Koresko98,Stapelfeldt03}. The geometrical situation we envision is depicted schematically in Figure \ref{diskcartoon}. While the available data do not allow a distinction between the two-disk scenario (Case A) and one in which the gas is located in the outer parts of a single inclined disk (Case B), the former scenario is geometrically
more likely. In either case, we are observing gas in the outer disk, and thus our analysis of the oxygen isotope ratios via CO is not predicated on the differentiation between the two geometries, although we note that Case B will minimize systematic differences between lines of sight at 2.3 and 4.7\,$\mu$m. 

The projected separation of the VV CrA components is 2.1\arcsec, corresponding
to 273 AU at the distance of 130 pc to the Corona Australis star-forming region \citep{Casey98, Neuhauser08}. The VV CrA system is
highly variable, as indicated by a K-band flux of $-$2.3 mag in 1987 \citep{Chelli95}, $0.3 - 0.0$ mag in 1993/95 \citep{Ageorges97,Ghez97} 
and 1.3 mag at the time of our observation in 2007. Due to the fact that the southern component has remained relatively constant
at K $\sim$ 6 mag, this corresponds to the IRC fading by $\sim$ 3 magnitudes in K in the past 20 years.  Given the geometry we propose in Figure \ref{diskcartoon} (Case A), this variability could be a result either of intrinsic variability of the continuum source, or to column density variations in the orbiting disk of the companion that move across our line of sight to the continuum source. It is unlikely to be due to temperature or composition changes on such short time scales.

The second source of this study is \object{Reipurth 50} [RE 50] (IRAS 05380-0728) (J2000: RA = 05 40 27.7, dec = $-$07 27 28), 
a luminous (250 $L_{\odot}$), embedded YSO in the Orion star-forming cloud at a distance of 470 pc. 
It illuminates a large, bright reflection nebula that appeared sometime between 1960 and 1970 
\citep{Reipurth86,Reipurth97}.  \cite{Strom93} found
that Reipurth 50 is an FU Ori type YSO, and it has been shown to be surrounded by 
a compact 0.1\,$M_{\odot}$ envelope \citep{Sandell01}, placing it in stage I \citep{Robitaille06} of the YSO evolutionary sequence.
It is likely that the absorbing gas is located in this envelope,
rather than in a disk. Thus, Reipurth 50 and VV CrA may provide a comparison between disk and envelope material, although larger samples of
objects are required to make any firm conclusions.

\section{Observations}
The observations of VV CrA and Reipurth 50 were taken as part of a large program to observe about 100 YSOs and protoplanetary disks with 
the newly-implemented Cryogenic Infrared Echelle Spectrograph (CRIRES) at the
Very Large Telescope (VLT) in Chile. 
CRIRES is an AO-assisted spectrometer that operates at very high resolving powers (for infrared spectroscopy) of $\lambda$/$\Delta$$\lambda$ $\approx$ 100\,000. 
The spectra of these sources allow for accurate determination of the CO column densities and isotopologue ratios. Table \ref{Obs_table} summarizes the observing parameters for VV CrA and Reipurth 50.

\begin{table}
\centering
\caption{Journal of Observations.}
\begin{tabular}{lccc}
\hline
\hline
Source/Obs. Date	&Spect. Range	&Int. time\tablenotemark{a}		& \\
(UT)	&($\mu$m)	&(minutes)	&S/N\tablenotemark{b} \\		
	\hline
VV CrA/2007-08-31	&$4.701 - 4.815$	&32	&260 \\
VV CrA/2007-08-31	&$2.338 - 2.396$	&23 &210 \\
Reipurth 50/2007-10-11	&$4.645 - 4.901$	&10-20	&290 \\
Reipurth 50/2007-10-17	&$2.338 - 2.397$	&20	&40\\
\hline
\end{tabular}
\tablenotetext{ a}{Integration is variable with wavelength, so values are approximate.}
\tablenotetext{ b}{Median values.}
\label{Obs_table}
\end{table}

\begin{figure*}
\includegraphics[width=17cm]{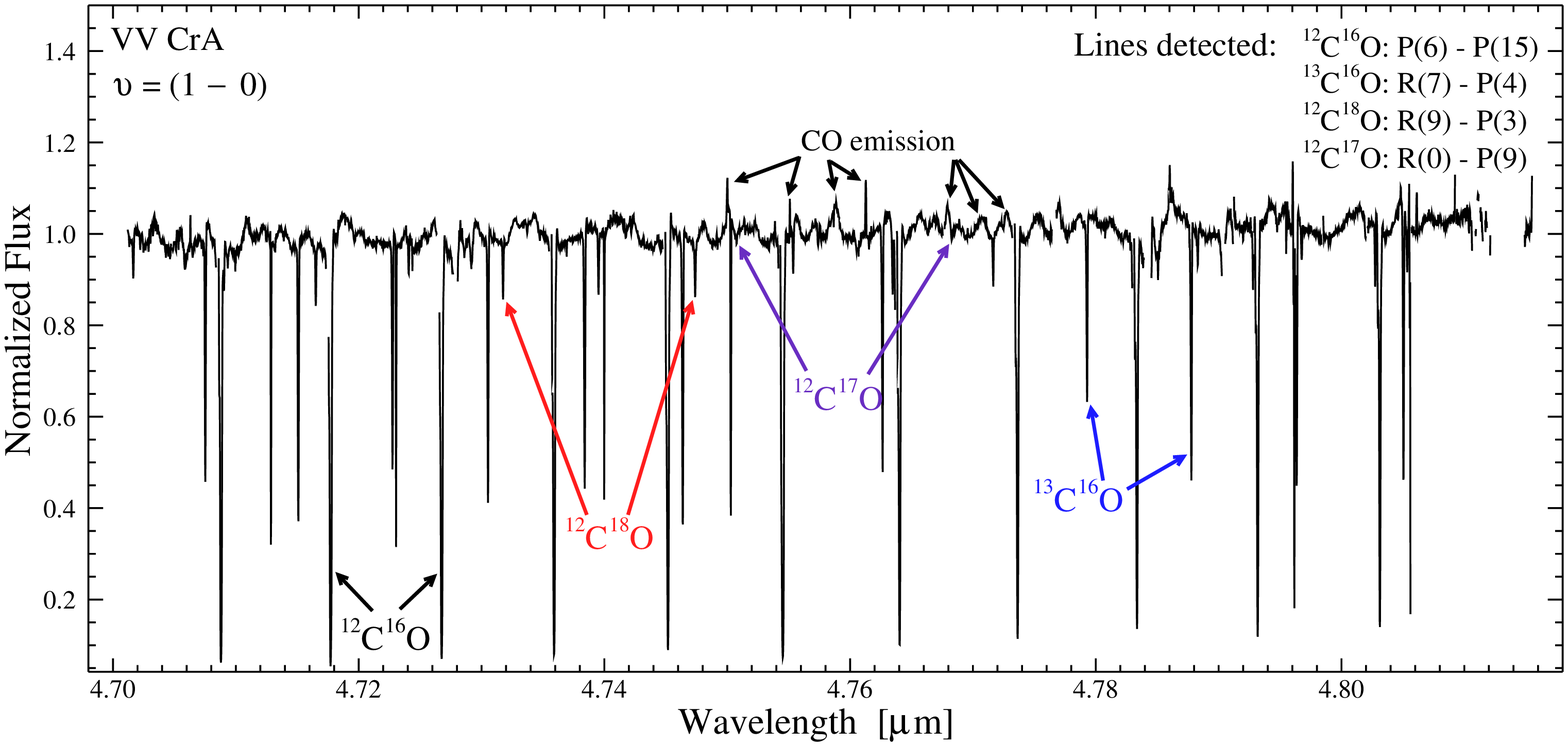}
\includegraphics[width=17cm]{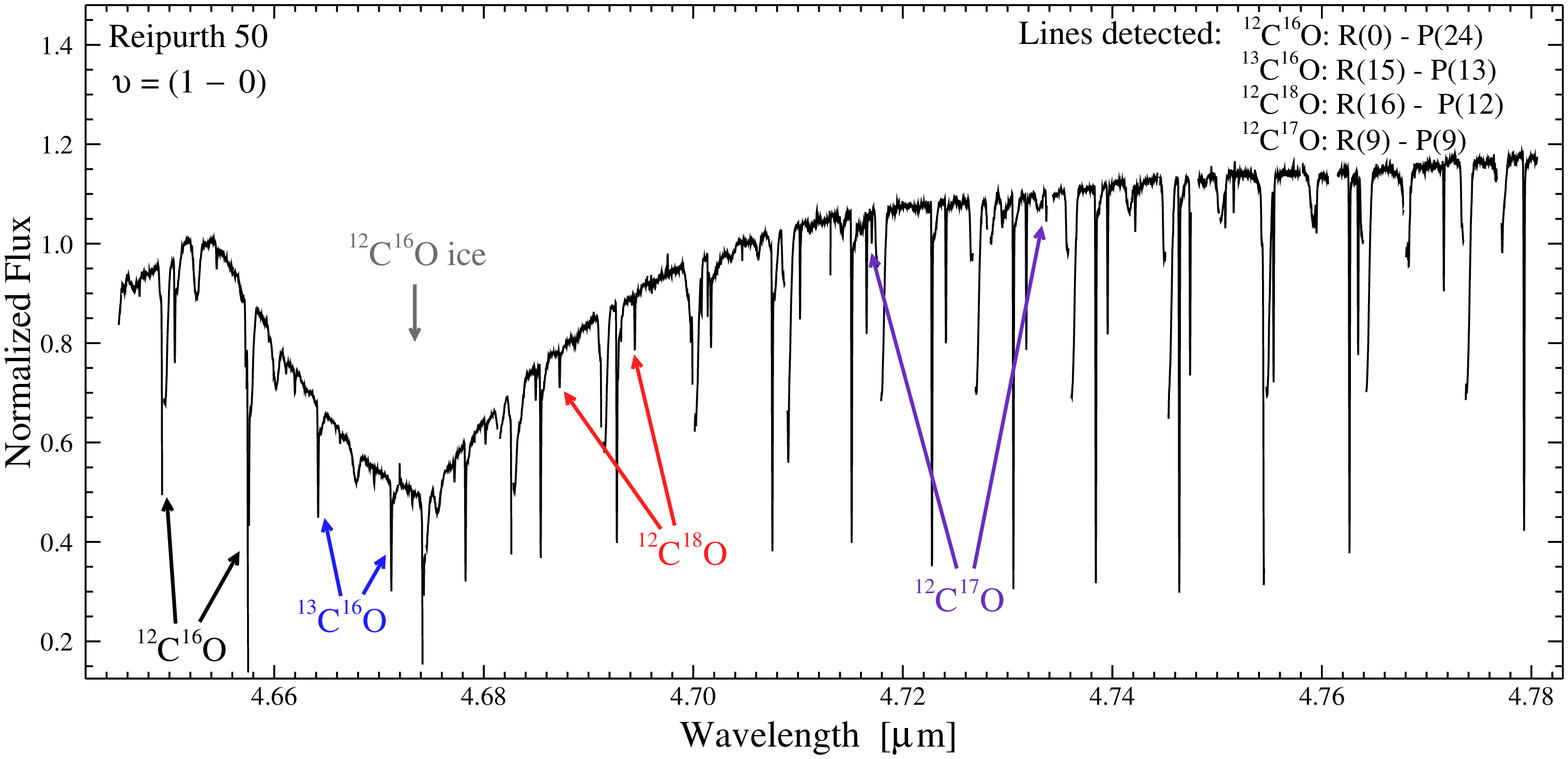}
\caption[]{Infrared absorption spectra of the CO fundamental ro-vibrational bands toward VV CrA and Reipurth 50. The fully observed spectral range is represented for VV CrA; for Reipurth 50, the portion with absorption lines used for the analysis is shown. Absorption lines are due to various CO isotopologues, representatives of which are indicated. 
Note that, in addition to the narrow absorption lines, VV CrA shows complex, low-level, broad-lined emission from hot CO gas, presumably located in the inner disk. The broad $\rm ^{12}C^{16}O$ ice feature in Reipurth 50 is also marked.}
\label{spectram}
\end{figure*} 

\begin{figure*}
\includegraphics[width=17cm]{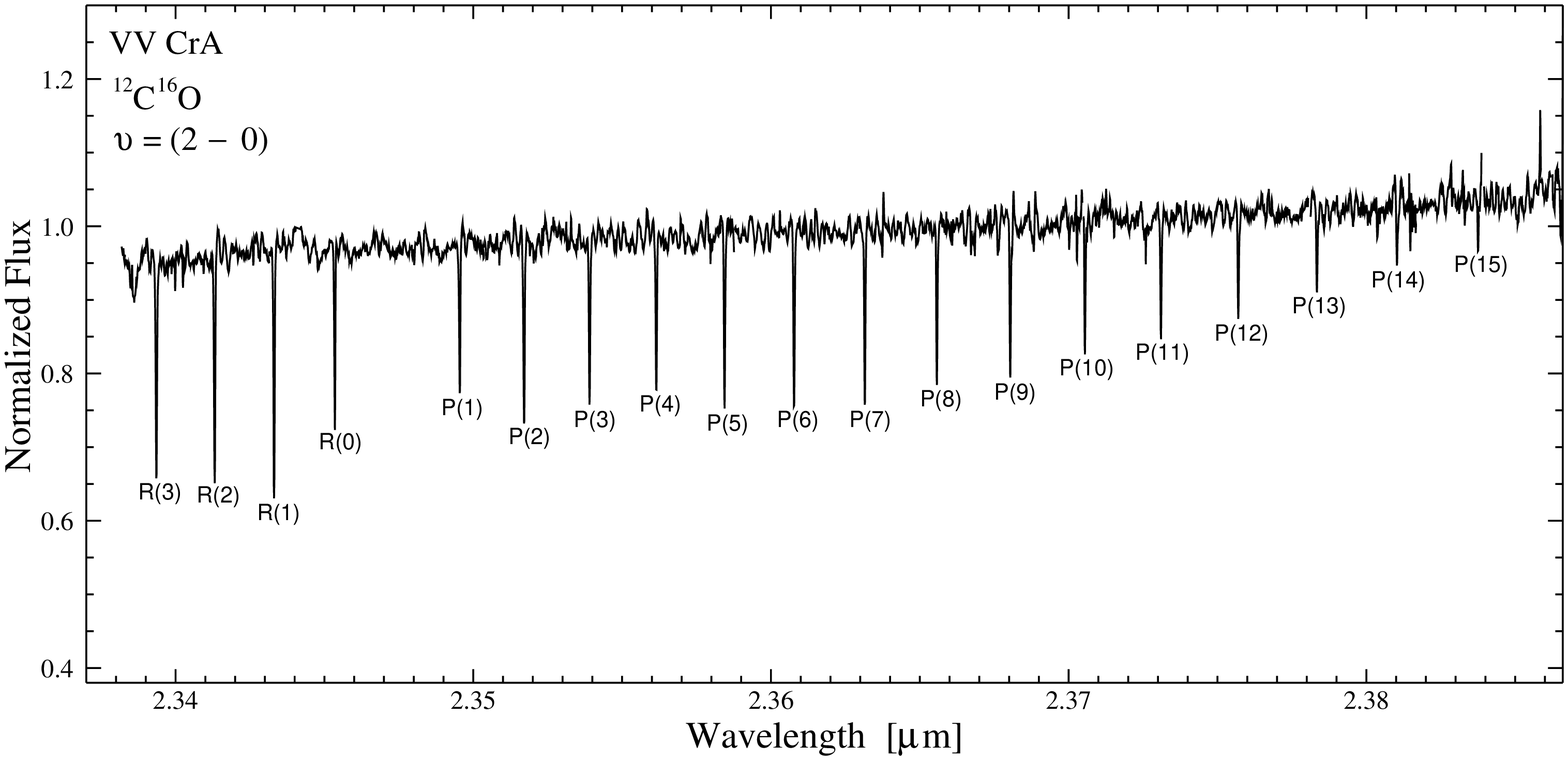}
\includegraphics[width=17cm]{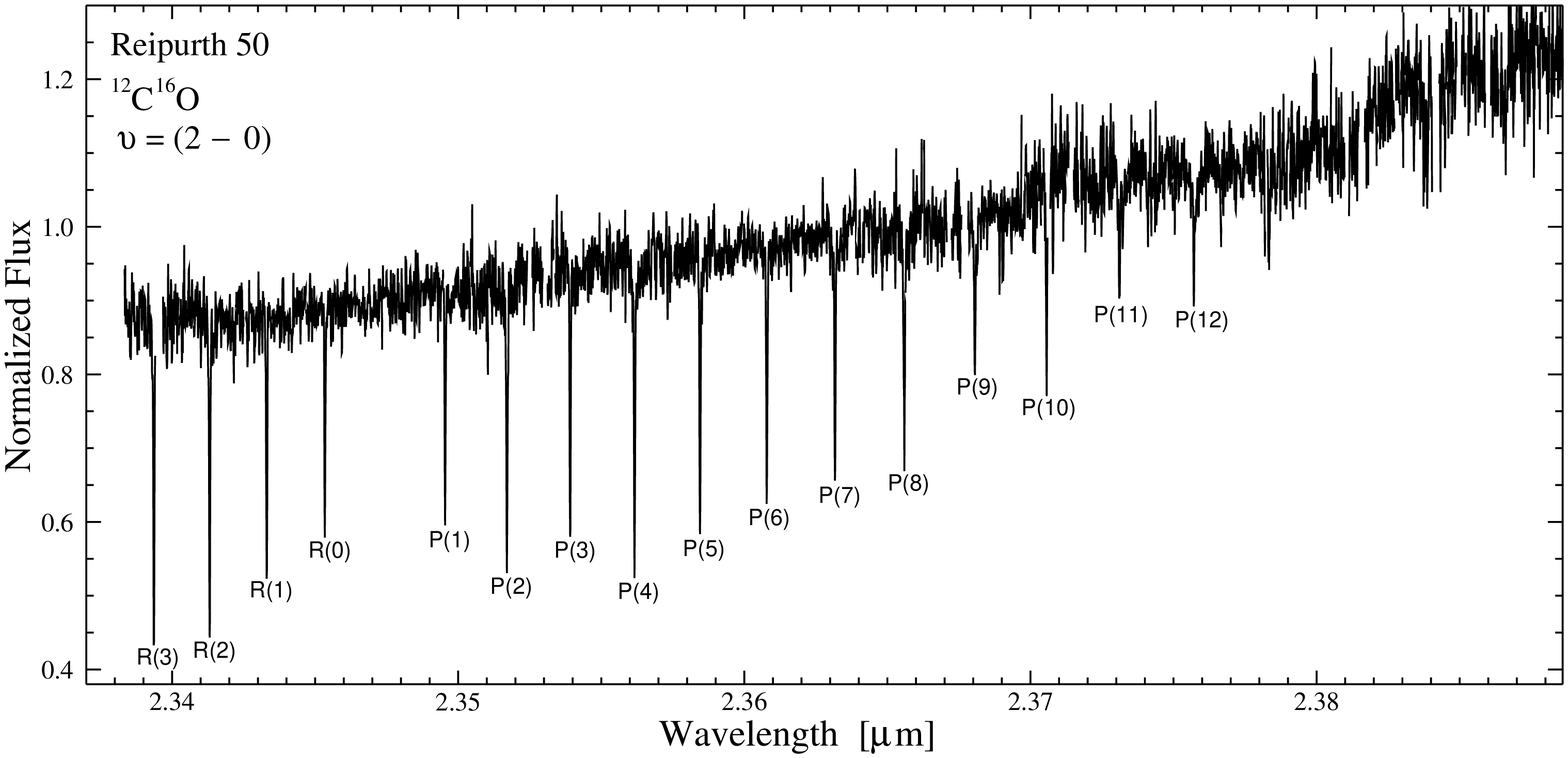}
\caption[]{Infrared absorption spectra of the CO overtone ro-vibrational band toward VV CrA [P(16)-P(17) not shown] and Reipurth 50. Detected $\rm ^{12}C^{16}O$ lines are marked.}
\label{spectrak}
\end{figure*} 

The spectrum of  VV CrA was observed on August 31, 2007 in both the M (4.7\,$\mu$m) and K (2.3\,$\mu$m) bands.  
Reipurth 50 was observed on October 11 and 17, 2007, respectively, and the observations from both dates were combined.  
CRIRES M-band spectra of VV CrA obtained in April 2007 showed variations in the depths of the CO lines 
of $\sim 10$ percent relative to the August data, illustrating the importance of obtaining nearly concurrent K- and M-band spectra. The spectra of the $v = (1-0)$ fundamental ro-vibrational bands for C$^{17}$O, C$^{18}$O, 
and $^{13}$CO and the $v = (2-0)$ overtone band of C$^{16}$O for VV CrA were obtained using the 0.2\arcsec \ 
slit, resulting
in a resolving power of $R = \lambda/\Delta\lambda \approx 95\,000$ (corresponding to 3.16 km s$^{-1}$), as measured on
unresolved telluric lines. Several settings were obtained for each source 
to probe a range of rotational levels for all the isotopologues, spanning $J = 0$ to at least $J = 9$. For $^{12}$CO, the $v = (2-0)$ transitions
were used because they have much smaller line strengths (by a factor of $\sim 130$) than the fundamental, and are thus expected to be optically thin (or nearly so). 

The data were reduced using standard procedures, including flat-field correction, adjustments to account for detector non-linearity and
linearization of the spectra in both the dispersion- and cross-dispersion directions. The spectra were wavelength calibrated using 
the telluric absorption lines referenced to an atmospheric model spectrum and transformed to the local standard of rest frame. 
Relative flux calibration was carried out by dividing the target sources by standard spectra of the stars HR 7236 (B9V) and HR 1666 (A3III). For more information on the CRIRES data processing, see \cite{2008ApJ...684.1323P}.

\section{Results and Data Analysis}
\label{Spectra}

The spectra of the fundamental and overtone CO
bands are shown in Figures \ref{spectram} and  \ref{spectrak}, respectively, illustrating the forest of narrow absorption lines
due to CO gas. While we are only interested in the narrow CO absorption lines, it is noteworthy that VV CrA also shows broad, complex
emission from CO, probably originating in the innermost ($R<1\,$AU) regions of the disk. Reipurth 50 does not exhibit an emission component, but the known absorption band from solid CO centered on 4.67\,$\mu$m can be seen. Given that the respective emission and ice features in these sources are much broader
than the absorption lines, their influence on derived gas CO column densities is minimal. In addition, the velocity shift of the center of the broad emission lines with respect to the narrow absorption lines are consistent with, and supportive of, our picture in which the absorption takes place in a different structure than the disk around the continuum source.

\begin{figure}
\vspace{0.8cm}
\includegraphics[width=4.5cm]{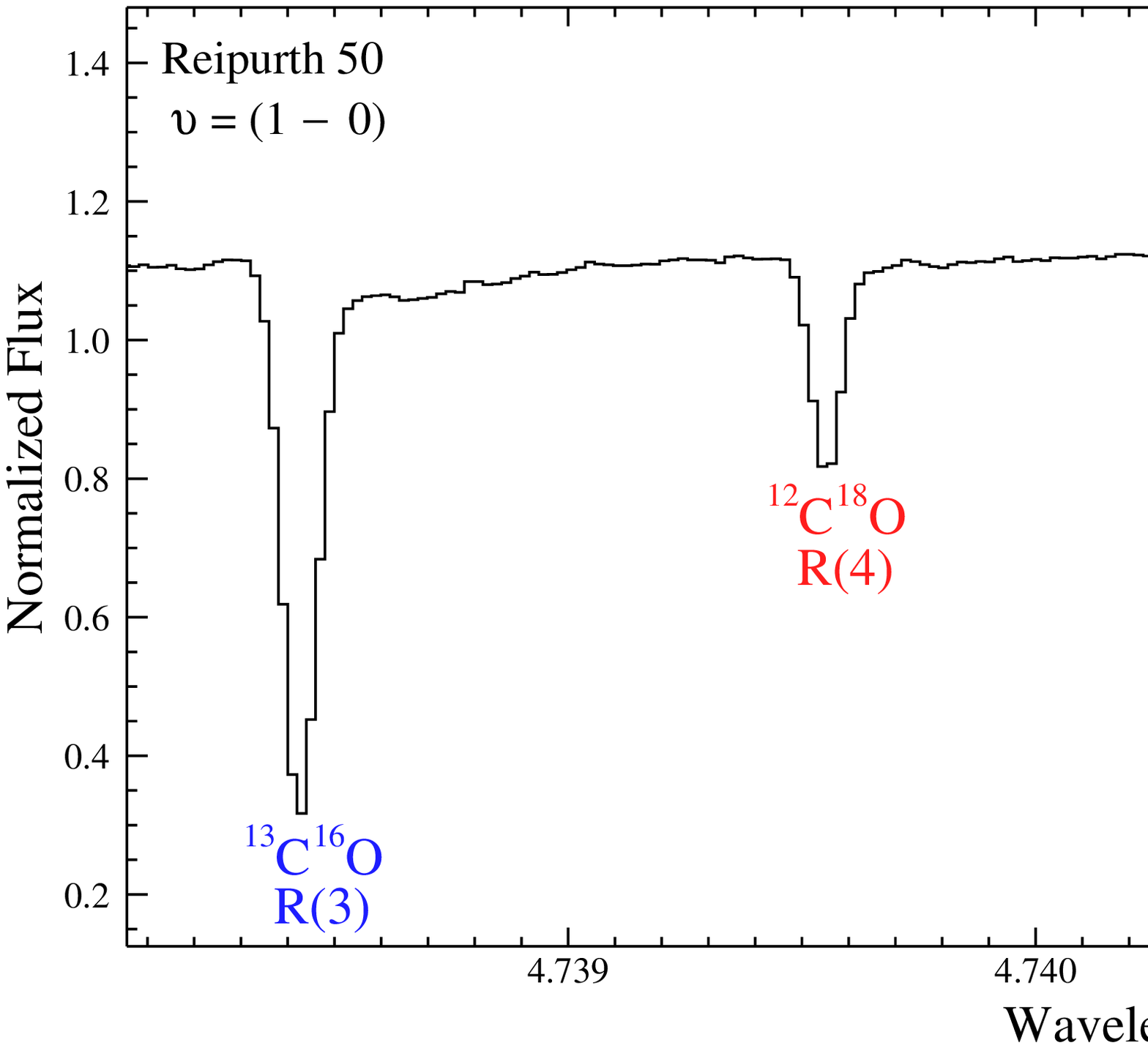}
\caption[]{Representative magnification of CO ro-vibrational lines for rare isotopologues in Reipurth 50.}
\label{detailspectram}
\end{figure}

Multiple lines from the four most abundant isotopologues of CO are detected in both sources, including the rare species C$^{17}$O (Figure \ref{detailspectram}).
$\rm C^{17}O$ has been observed in the ultraviolet toward the X Persei molecular 
cloud \citep{Sheffer2002-L171} and the $P(1)$ $\rm C^{17}O$ ro-vibrational line has been recently reported in the T Tauri star SR21 in infrared emission \citep{2008ApJ...684.1323P}. While the latter observation would potentially allow a window into measuring isotope ratios in the inner $<$1 AU regions of disks and unambiguous determination of the location of the gas, radiative transfer modeling for emission lines is more complicated than for pencil-beam absorption lines.  We emphasize that our observations of all four CO isotopologues in infrared absorption probe a single sight line, therefore obviating the need for complex modeling required for emission lines.

\begin{figure*}
\includegraphics[width=18cm]{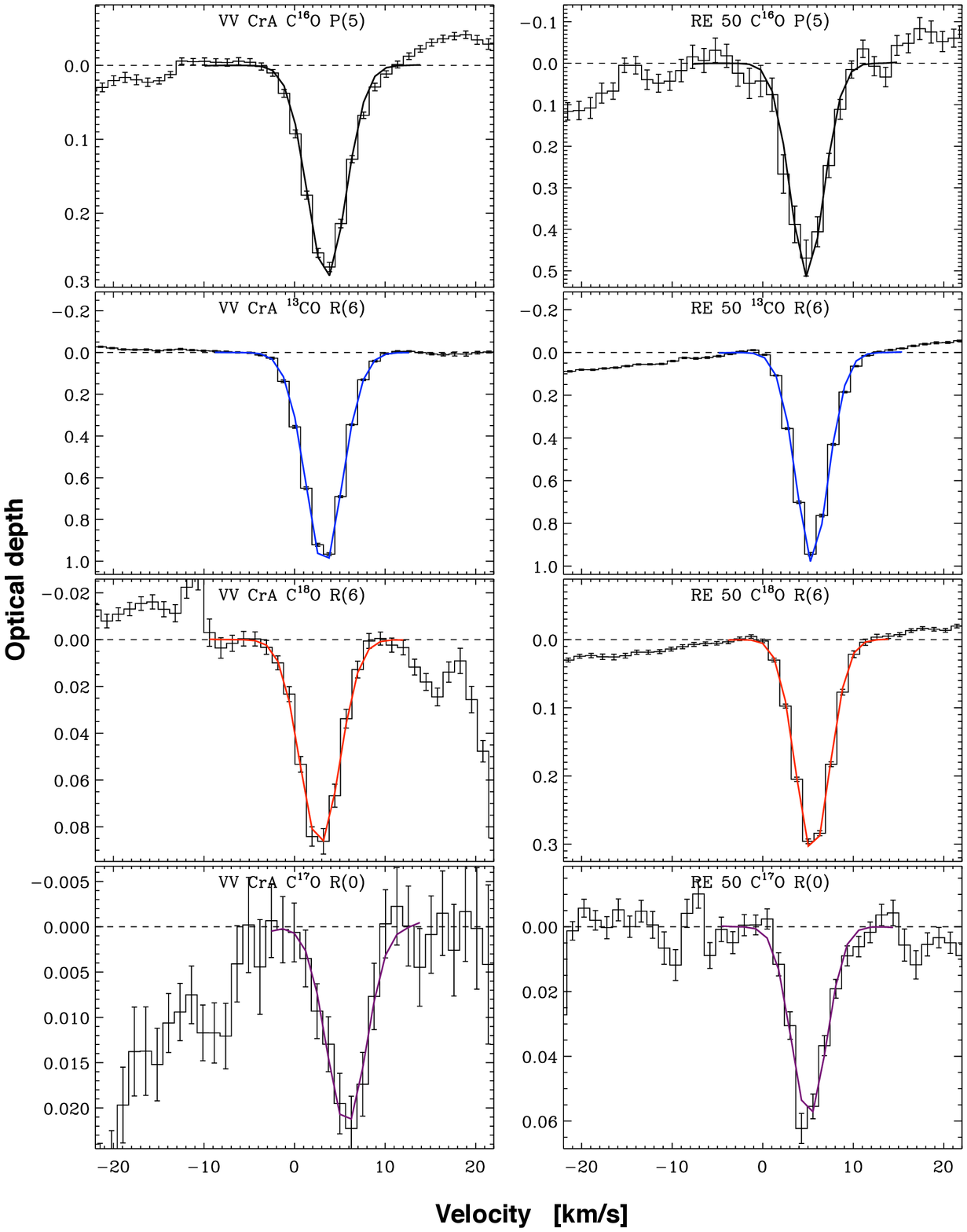}
\caption[]{Representative fits of CO isotopologues for VV CrA and Reipurth 50. Fits to $\rm C^{16}O$ lines are to the overtone band. All other fits are to the fundamental spectra. The fits are shown here normalized to the linear best-fit baseline. Error bars are 1$\sigma$/velocity channel. Statistical errors do not include telluric residuals (see text).}
\label{specfits}
\end{figure*}

\subsection{Derivation of column densities}

In order to determine the total column density for each isotopologue, the column density of 
CO in each rotational state must be measured; the total column density of CO can
then be calculated by summing the measured column densities in the observed ro-vibrational levels and assuming that the remainder are populated according to a Boltzmann distribution. 
If lines are spectrally unresolved, accurate abundance derivations are complicated by the fact that it is difficult to determine optical depths, necessitating a 
curve-of-growth analysis. Given the high spectral resolution of CRIRES, the data reported in this study are not complicated by these factors, and we are able to distinguish
intrinsic line broadening from that of the instrument. Figure \ref{detailspectram} is a representative section of
the fundamental CO band in Reipurth 50, illustrating the quality of the data. The CO lines were confirmed to be spectrally resolved by comparing them to adjacent
unresolved telluric lines; the total CO line widths are slightly, but significantly, broader than the instrumental resolution of 3.2\,$\rm km\,s^{-1}$ exhibited by the telluric lines. Column densities were obtained from the optical depths at line center $(\tau _{\circ})$ derived from the best-fit profiles to each line. Tables \ref{vvcra_table} and \ref{re50_table} show the optical depths at line center and Doppler shifts $\rm (V_{LSR})$ for each observed line used in the derivation of isotopologue column densities for VV CrA and Reipurth 50, respectively. 

\begin{table}
\tiny
\centering
\caption{Optical depths and Doppler shifts for observed CO isotopologue ro-vibrational lines used in deriving column densities in VV CrA. Uncertainties are $1\sigma$ derived from the line fits.}
\begin{tabular}{c l c c}
\hline
\hline
\multicolumn{4}{{c}}{VV CrA} \\ \hline
&		&Optical depth 		&Doppler shift \\
Isotopologue		&Line ID	&$(\tau _{\circ})$	& $\rm  (V_{LSR}, km \ s^{-1})$ \\
\hline
{$^{12}$C$^{16}$O} 	&$(2, 0)\ R(3)$ 		&$\rm0.475\pm0.005$	&$\rm4.16\pm0.03$\\
{$^{12}$C$^{16}$O}		&$(2, 0)\ R(2)$		&$\rm0.535\pm0.006$	&$\rm4.77\pm0.02$\\
{$^{12}$C$^{16}$O}		&$(2, 0)\ R(1)$		&$\rm0.581\pm0.006$	&$\rm5.08\pm0.02$\\
{$^{12}$C$^{16}$O}		&$(2, 0)\ R(0)$		&$\rm0.371\pm0.005$	&$\rm5.61\pm0.03$\\
{$^{12}$C$^{16}$O}		&$(2, 0)\ P(1)$		&$\rm0.309\pm0.005$	&$\rm5.41\pm0.04$\\
{$^{12}$C$^{16}$O}		&$(2, 0)\ P(2)$		&$\rm0.397\pm0.004$	&$\rm5.01\pm0.03$\\
{$^{12}$C$^{16}$O}		&$(2, 0)\ P(3)$		&$\rm0.337\pm0.005$	&$\rm4.21\pm0.03$\\
{$^{12}$C$^{16}$O}		&$(2, 0)\ P(4)$		&$\rm0.319\pm0.005$	&$\rm3.57\pm0.04$\\
{$^{12}$C$^{16}$O}		&$(2, 0)\ P(5)$		&$\rm0.371\pm0.005$	&$\rm3.55\pm0.04$\\
{$^{12}$C$^{16}$O}		&$(2, 0)\ P(6)$		&$\rm0.381\pm0.004$	&$\rm3.67\pm0.03$\\
{$^{12}$C$^{16}$O}		&$(2, 0)\ P(7)$		&$\rm0.360\pm0.005$	&$\rm3.62\pm0.03$\\
{$^{12}$C$^{16}$O}		&$(2, 0)\ P(8)$		&$\rm0.310\pm0.005$	&$\rm3.53\pm0.04$\\
{$^{12}$C$^{16}$O}		&$(2, 0)\ P(9)$		&$\rm0.287\pm0.005$	&$\rm3.43\pm0.04$\\
{$^{12}$C$^{16}$O}		&$(2, 0)\ P(10)$	&$\rm0.255\pm0.009$	&$\rm3.46\pm0.08$\\
{$^{12}$C$^{16}$O}		&$(2, 0)\ P(11)$	&$\rm0.231\pm0.005$	&$\rm3.78\pm0.05$\\
{$^{12}$C$^{16}$O}		&$(2, 0)\ P(12)$	&$\rm0.206\pm0.004$	&$\rm4.08\pm0.05$\\
{$^{12}$C$^{16}$O}		&$(2, 0)\ P(13)$	&$\rm0.154\pm0.005$	&$\rm3.99\pm0.08$\\
{$^{12}$C$^{16}$O}		&$(2, 0)\ P(14)$	&$\rm0.136\pm0.005$	&$\rm4.78\pm0.09$\\
{$^{12}$C$^{16}$O}		&$(2, 0)\ P(15)$	&$\rm0.088\pm0.007$	&$\rm3.94\pm0.16$\\
{$^{12}$C$^{16}$O}		&$(2, 0)\ P(16)$	&$\rm0.093\pm0.007$	&$\rm3.72\pm0.17$\\
{$^{12}$C$^{16}$O}		&$(2, 0)\ P(17)$	&$\rm0.064\pm0.005$	&$\rm4.28\pm0.21$\\ 
$^{13}$C$^{16}$O 		&$(1, 0)\ R(7)$		&$\rm1.038\pm0.010$	&$\rm3.17\pm0.02$ \\
$^{13}$C$^{16}$O		&$(1, 0)\ R(6)$		&$\rm1.381\pm0.006$	&$\rm3.25\pm0.01$ \\
$^{13}$C$^{16}$O		&$(1, 0)\ R(5)$		&$\rm0.880\pm0.007$	&$\rm3.08\pm0.01$ \\
$^{13}$C$^{16}$O		&$(1, 0)\ R(4)$		&$\rm1.159\pm0.007$	&$\rm3.38\pm0.01$ \\
$^{13}$C$^{16}$O		&$(1, 0)\ R(3)$		&$\rm1.104\pm0.008$	&$\rm3.30\pm0.01$ \\
$^{13}$C$^{16}$O		&$(1, 0)\ R(2)$		&$\rm1.450\pm0.008$	&$\rm4.00\pm0.01$ \\
$^{13}$C$^{16}$O		&$(1, 0)\ R(0)$		&$\rm0.921\pm0.005$	&$\rm5.09\pm0.01$ \\
$^{13}$C$^{16}$O		&$(1, 0)\ P(1)$		&$\rm0.611\pm0.006$	&$\rm4.23\pm0.02$ \\
$^{13}$C$^{16}$O		&$(1, 0)\ P(2)$		&$\rm1.025\pm0.010$	&$\rm4.40\pm0.02$ \\
$^{13}$C$^{16}$O		&$(1, 0)\ P(4)$		&$\rm1.091\pm0.007$	&$\rm3.23\pm0.01$ \\ 
$^{12}$C$^{18}$O		&$(1, 0)\ R(9)$		&$\rm0.124\pm0.003$	&$\rm2.75\pm0.06$ \\
$^{12}$C$^{18}$O		&$(1, 0)\ R(7)$		&$\rm0.180\pm0.004$	&$\rm3.32\pm0.07$ \\
$^{12}$C$^{18}$O		&$(1, 0)\ R(6)$		&$\rm0.112\pm0.004$	&$\rm2.79\pm0.08$ \\
$^{12}$C$^{18}$O		&$(1, 0)\ R(5)$		&$\rm0.179\pm0.004$	&$\rm3.38\pm0.05$ \\
$^{12}$C$^{18}$O		&$(1, 0)\ R(4)$		&$\rm0.186\pm0.003$	&$\rm3.39\pm0.04$ \\
$^{12}$C$^{18}$O		&$(1, 0)\ R(3)$		&$\rm0.164\pm0.006$	&$\rm3.20\pm0.09$ \\
$^{12}$C$^{18}$O		&$(1, 0)\ R(2)$		&$\rm0.136\pm0.003$	&$\rm3.81\pm0.06$ \\
$^{12}$C$^{18}$O		&$(1, 0)\ R(1)$		&$\rm0.180\pm0.005$	&$\rm5.02\pm0.06$ \\
$^{12}$C$^{18}$O		&$(1, 0)\ R(0)$		&$\rm0.141\pm0.004$	&$\rm5.79\pm0.07$ \\
$^{12}$C$^{18}$O		&$(1, 0)\ P(1)$		&$\rm0.102\pm0.004$	&$\rm5.21\pm0.09$ \\
$^{12}$C$^{18}$O		&$(1, 0)\ P(2)$		&$\rm0.103\pm0.004$	&$\rm3.53\pm0.09$ \\
$^{12}$C$^{18}$O		&$(1, 0)\ P(3)$		&$\rm0.133\pm0.006$	&$\rm3.20\pm0.11$ \\	
$^{12}$C$^{17}$O		&$(1, 0)\ R(0)$		&$\rm0.028\pm0.004$	&$\rm5.72\pm0.31$ \\
$^{12}$C$^{17}$O		&$(1, 0)\ P(1)$		&$\rm0.027\pm0.003$	&$\rm5.72\pm0.26$ \\
$^{12}$C$^{17}$O		&$(1, 0)\ P(2)$		&$\rm0.039\pm0.004$	&$\rm4.72\pm0.25$ \\
$^{12}$C$^{17}$O		&$(1, 0)\ P(3)$		&$\rm0.024\pm0.004$	&$\rm3.86\pm0.40$ \\
$^{12}$C$^{17}$O		&$(1, 0)\ P(5)$		&$\rm0.045\pm0.003$	&$\rm2.82\pm0.18$ \\
$^{12}$C$^{17}$O		&$(1, 0)\ P(6)$		&$\rm0.034\pm0.005$	&$\rm3.83\pm0.36$ \\
$^{12}$C$^{17}$O		&$(1, 0)\ P(8)$		&$\rm0.022\pm0.003$      &$\rm3.10\pm0.38$ \\
$^{12}$C$^{17}$O		&$(1, 0)\ P(9)$		&$\rm0.034\pm0.004$	&$\rm4.27\pm0.32$ \\
\end{tabular}
\label{vvcra_table}
\end{table}

\begin{table}
\tiny
\centering
\caption{Optical depths and Doppler shifts for observed CO isotopologue ro-vibrational lines used in deriving column densities in Reipurth 50. Uncertainties are $1\sigma$ derived from the line fits.}
\begin{tabular}{c l c c}
\hline
\hline
\multicolumn{4}{{c}}{Reipurth 50} \\ \hline
&		&Optical depth 		&Doppler shift \\
Isotopologue		&Line ID	&$(\tau _{\circ})$	& $\rm  (V_{LSR}, km \ s^{-1})$ \\
\hline
$^{12}$C$^{16}$O	&{$(2, 0)\ R(3)$} 	&$\rm1.013\pm0.05$	&$\rm5.31\pm0.09$ \\
$^{12}$C$^{16}$O	&{$(2, 0)\ R(2)$} 	&$\rm1.035\pm0.04$	&$\rm4.91\pm0.08$ \\
$^{12}$C$^{16}$O	&{$(2, 0)\ R(1)$} 	&$\rm0.847\pm0.04$	&$\rm4.94\pm0.10$ \\
$^{12}$C$^{16}$O	&{$(2, 0)\ R(0)$} 	&$\rm0.611\pm0.03$	&$\rm4.35\pm0.11$ \\
$^{12}$C$^{16}$O	&{$(2, 0)\ P(1)$} 	&$\rm0.619\pm0.04$	&$\rm4.85\pm0.12$ \\
$^{12}$C$^{16}$O	&{$(2, 0)\ P(2)$} 	&$\rm0.775\pm0.05$	&$\rm4.69\pm0.13$ \\
$^{12}$C$^{16}$O	&{$(2, 0)\ P(3)$} 	&$\rm0.799\pm0.04$	&$\rm5.08\pm0.09$ \\
$^{12}$C$^{16}$O	&{$(2, 0)\ P(4)$} 	&$\rm0.796\pm0.04$	&$\rm5.36\pm0.10$ \\
$^{12}$C$^{16}$O	&{$(2, 0)\ P(5)$} 	&$\rm0.752\pm0.05$	&$\rm4.89\pm0.13$ \\
$^{12}$C$^{16}$O	&{$(2, 0)\ P(6)$} 	&$\rm0.658\pm0.03$	&$\rm4.90\pm0.11$ \\
$^{12}$C$^{16}$O	&{$(2, 0)\ P(7)$} 	&$\rm0.609\pm0.03$	&$\rm5.67\pm0.12$ \\
$^{12}$C$^{16}$O	&{$(2, 0)\ P(8)$} 	&$\rm0.486\pm0.04$	&$\rm5.79\pm0.15$ \\
$^{12}$C$^{16}$O	&{$(2, 0)\ P(9)$} 	&$\rm0.291\pm0.03$	&$\rm5.81\pm0.23$ \\
$^{12}$C$^{16}$O	&{$(2, 0)\ P(11)$} 	&$\rm0.220\pm0.03$	&$\rm5.26\pm0.30$ \\
$^{12}$C$^{16}$O	&{$(2, 0)\ P(12)$} 	&$\rm0.189\pm0.03$	&$\rm5.58\pm0.33$ \\
$^{13}$C$^{16}$O	&$(1, 0)\ R(15)$		&$\rm0.314\pm0.004$	&$\rm5.59\pm0.03$ \\
$^{13}$C$^{16}$O	&$(1, 0)\ R(13)$		&$\rm0.558\pm0.006$	&$\rm5.59\pm0.02$ \\
$^{13}$C$^{16}$O	&$(1, 0)\ R(12)$		&$\rm0.746\pm0.010$	&$\rm5.66\pm0.03$ \\
$^{13}$C$^{16}$O	&$(1, 0)\ R(11)$		&$\rm0.881\pm0.009$	&$\rm5.84\pm0.02$ \\
$^{13}$C$^{16}$O	&$(1, 0)\ R(10)$		&$\rm1.044\pm0.008$	&$\rm5.63\pm0.01$ \\
$^{13}$C$^{16}$O	&$(1, 0)\ R(7)$			&$\rm1.451\pm0.012$	&$\rm5.43\pm0.01$ \\
$^{13}$C$^{16}$O	&$(1, 0)\ R(6)$			&$\rm1.510\pm0.006$	&$\rm5.44\pm0.01$ \\
$^{13}$C$^{16}$O	&$(1, 0)\ R(5)$			&$\rm1.877\pm0.007$	&$\rm5.13\pm0.00$ \\
$^{13}$C$^{16}$O	&$(1, 0)\ R(4)$			&$\rm1.955\pm0.007$	&$\rm5.26\pm0.00$ \\
$^{13}$C$^{16}$O	&$(1, 0)\ R(3)$			&$\rm2.099\pm0.008$	&$\rm5.26\pm0.00$ \\
$^{13}$C$^{16}$O	&$(1, 0)\ R(2)$			&$\rm2.449\pm0.011$	&$\rm4.94\pm0.01$ \\
$^{13}$C$^{16}$O	&$(1, 0)\ R(0)$			&$\rm1.840\pm0.006$	&$\rm4.74\pm0.00$ \\
$^{13}$C$^{16}$O	&$(1, 0)\ P(1)$			&$\rm1.612\pm0.005$	&$\rm5.17\pm0.00$ \\
$^{13}$C$^{16}$O	&$(1, 0)\ P(4)$			&$\rm1.978\pm0.012$	&$\rm5.08\pm0.01$ \\
$^{13}$C$^{16}$O	&$(1, 0)\ P(9)$			&$\rm1.244\pm0.008$	&$\rm3.61\pm0.01$ \\
$^{13}$C$^{16}$O	&$(1, 0)\ P(10)$		&$\rm1.087\pm0.008$	&$\rm5.56\pm0.01$ \\
$^{13}$C$^{16}$O	&$(1, 0)\ P(11)$		&$\rm0.897\pm0.009$	&$\rm6.09\pm0.02$ \\
$^{13}$C$^{16}$O	&$(1, 0)\ P(13)$		&$\rm0.611\pm0.007$	&$\rm4.35\pm0.02$ \\
$^{12}$C$^{18}$O	&$(1, 0)\ R(15)$	&$\rm0.025\pm0.003$	&$\rm5.56\pm0.30$ \\
$^{12}$C$^{18}$O	&$(1, 0)\ R(14)$	&$\rm0.051\pm0.004$	&$\rm5.67\pm0.19$ \\
$^{12}$C$^{18}$O	&$(1, 0)\ R(13)$	&$\rm0.084\pm0.006$	&$\rm6.02\pm0.18$ \\
$^{12}$C$^{18}$O	&$(1, 0)\ R(12)$	&$\rm0.101\pm0.005$	&$\rm6.05\pm0.12$ \\
$^{12}$C$^{18}$O	&$(1, 0)\ R(11)$	&$\rm0.142\pm0.004$	&$\rm5.66\pm0.06$ \\
$^{12}$C$^{18}$O	&$(1, 0)\ R(10)$	&$\rm0.196\pm0.003$	&$\rm5.44\pm0.04$ \\
$^{12}$C$^{18}$O	&$(1, 0)\ R(9)$		&$\rm0.239\pm0.004$	&$\rm5.40\pm0.04$ \\
$^{12}$C$^{18}$O	&$(1, 0)\ R(7)$		&$\rm0.372\pm0.003$	&$\rm5.55\pm0.02$ \\
$^{12}$C$^{18}$O	&$(1, 0)\ R(6)$		&$\rm0.450\pm0.003$	&$\rm5.52\pm0.01$ \\
$^{12}$C$^{18}$O	&$(1, 0)\ R(5)$		&$\rm0.485\pm0.003$	&$\rm5.78\pm0.01$ \\
$^{12}$C$^{18}$O	&$(1, 0)\ R(4)$		&$\rm0.457\pm0.003$	&$\rm5.21\pm0.02$ \\
$^{12}$C$^{18}$O	&$(1, 0)\ R(2)$		&$\rm0.651\pm0.003$	&$\rm5.08\pm0.01$ \\
$^{12}$C$^{18}$O	&$(1, 0)\ R(1)$		&$\rm0.568\pm0.004$	&$\rm5.13\pm0.01$ \\
$^{12}$C$^{18}$O	&$(1, 0)\ R(0)$		&$\rm0.348\pm0.002$	&$\rm5.15\pm0.01$ \\
$^{12}$C$^{18}$O	&$(1, 0)\ P(2)$		&$\rm0.456\pm0.004$	&$\rm5.15\pm0.02$ \\
$^{12}$C$^{18}$O	&$(1, 0)\ P(3)$		&$\rm0.516\pm0.005$	&$\rm5.21\pm0.02$ \\
$^{12}$C$^{18}$O	&$(1, 0)\ P(8)$		&$\rm0.312\pm0.004$	&$\rm3.90\pm0.03$ \\
$^{12}$C$^{18}$O	&$(1, 0)\ P(9)$		&$\rm0.238\pm0.006$	&$\rm5.57\pm0.06$ \\
$^{12}$C$^{18}$O	&$(1, 0)\ P(10)$	&$\rm0.193\pm0.004$	&$\rm5.98\pm0.05$ \\
$^{12}$C$^{17}$O	&$(1, 0)\ R(9)$		&$\rm0.049\pm0.003$	&$\rm5.36\pm0.15$ \\
$^{12}$C$^{17}$O	&$(1, 0)\ R(8)$		&$\rm0.068\pm0.003$	&$\rm5.29\pm0.10$ \\
$^{12}$C$^{17}$O	&$(1, 0)\ R(7)$		&$\rm0.082\pm0.004$	&$\rm5.68\pm0.11$ \\
$^{12}$C$^{17}$O	&$(1, 0)\ R(6)$		&$\rm0.097\pm0.006$	&$\rm5.78\pm0.13$ \\
$^{12}$C$^{17}$O	&$(1, 0)\ R(5)$		&$\rm0.114\pm0.006$	&$\rm5.87\pm0.11$ \\
$^{12}$C$^{17}$O	&$(1, 0)\ R(4)$		&$\rm0.128\pm0.004$	&$\rm5.27\pm0.07$ \\
$^{12}$C$^{17}$O	&$(1, 0)\ R(2)$		&$\rm0.132\pm0.003$	&$\rm4.80\pm0.05$ \\
$^{12}$C$^{17}$O	&$(1, 0)\ R(0)$		&$\rm0.082\pm0.003$	&$\rm5.10\pm0.08$ \\
$^{12}$C$^{17}$O	&$(1, 0)\ P(2)$		&$\rm0.115\pm0.003$	&$\rm5.08\pm0.06$ \\
$^{12}$C$^{17}$O	&$(1, 0)\ P(3)$		&$\rm0.105\pm0.002$	&$\rm5.25\pm0.05$ \\
$^{12}$C$^{17}$O	&$(1, 0)\ P(4)$		&$\rm0.097\pm0.003$	&$\rm5.36\pm0.06$ \\
$^{12}$C$^{17}$O	&$(1, 0)\ P(5)$		&$\rm0.106\pm0.003$	&$\rm5.48\pm0.06$ \\
$^{12}$C$^{17}$O	&$(1, 0)\ P(8)$		&$\rm0.054\pm0.005$	&$\rm6.07\pm0.19$ \\
$^{12}$C$^{17}$O	&$(1, 0)\ P(9)$		&$\rm0.041\pm0.003$	&$\rm5.47\pm0.17$ \\
\end{tabular}
\label{re50_table}
\end{table}

The fundamental $v=(1-0)$ $\rm C^{16}O$ lines are optically thick. Therefore, the $\rm C^{16}O$ column densities were measured using 
$v = (2-0)$ transitions in order to use only lines that are as optically thin as possible.  All line fits are consistent
in having a single, constant width, within error, as expected for optically thin lines. That width is substantially larger than the thermal linewidth (see below), and so is likely due to some combination of local turbulent motions and radial velocity gradients transverse to the absorbing column. 

We fit the frequency-dependent flux, $I_\nu$, for any given transition using 

\begin{equation}
I_\nu   = I_c \exp ( - s\;\phi _\nu  N_{vJ} ),
\label{eq:intensity}
\end{equation}
where $I_c$ is the flux of the continuum and $\phi_\nu$ is the intrinsic line profile (normalized to unity). The intrinsic
line profile is assumed to
be Gaussian with a full width at half maximum (FWHM) = $\gamma_{\rm intrinsic}$. $N_{vJ}$ is the column density of the lower level of the transition, having rotational and vibrational quantum numbers $J$ and $v$ ($v$ = 0 for all
observations presented here), $s$ is the integrated cross section, and $\tau_{\circ} \equiv sN_{vJ}\phi_{\circ}$. 

The integrated cross section ($\rm cm^2 s^{-1}$) (uncorrected for stimulated emission) is   
\begin{equation}
s = \frac{{\pi e^2 }}{{m_e c}}f_{v'J' \leftarrow  vJ},
\label{eq:cross_section}
\end{equation}
\citep{Spitzer1978-Physical}, where $f_{v'J' \leftarrow vJ}$ is the oscillator strength for absorption from level $v$, $J$ to level $v'$,$ J'$, and is related to the Einstein $A$ 
coefficient by

\begin{equation}
A_{v'J' \rightarrow vJ} = \frac{8\nu^2\pi^2e^2}{m_ec^3} \frac{g_J}{g_{J'}} f_{v'J' \leftarrow vJ}
\label{eq:A-f}
\end{equation}
Einstein $A$s were taken from the HITRAN database \citep{HITRAN}, which uses the values
of \cite{Goorvitch94A,Goorvitch94B}. \cite{Ayres06} compared different sets of molecular parameters for 
the CO fundamental band and found our adopted values to be consistent with other works.  

\begin{figure}
\centering
\vspace{0.2cm}
\includegraphics[width=8.5cm]{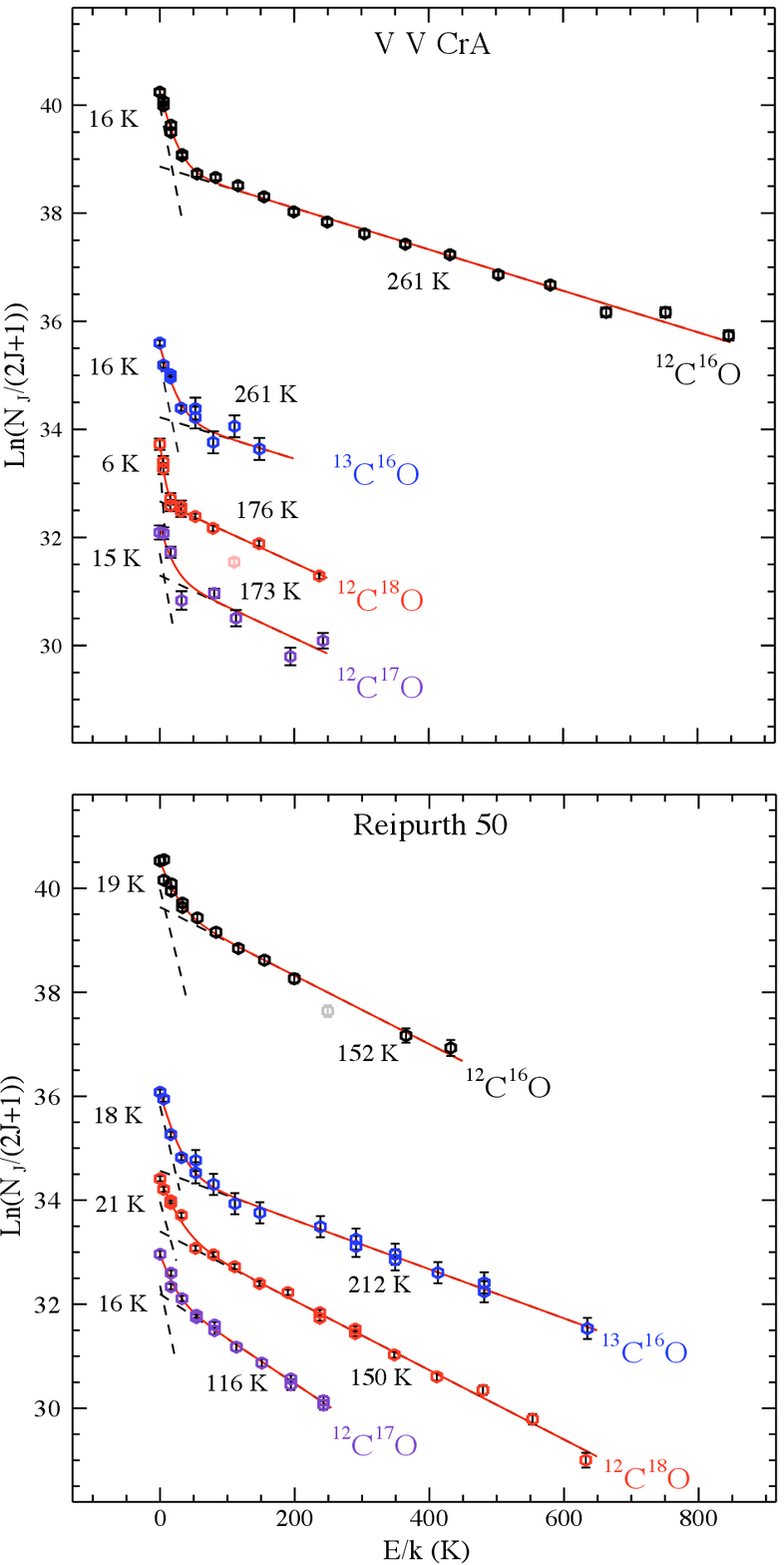}
\caption[]{Rotational excitation diagrams for CO isotopologues in VV CrA and Reipurth 50. Simultaneous fits to the two-temperature model and derived high and low temperatures for each isotopologue are indicated by the red curves and dashed lines, respectively. Error bars are $1\sigma$ propagated from the Gaussian fits. $E_J$ is the energy of the $J$th 
rotational state above the ground rotational state and $k$ is the Boltzmann constant. Faded symbols have been excluded from the fits.}
\label{rotplot}
\end{figure}

Because the intrinsic line widths are close to the instrumental resolution, the $I_\nu$ given by (\ref{eq:intensity}) is 
convolved with the instrumental line profile modeled by a Gaussian with FWHM = 3.2\,$\rm km\,s^{-1}$ prior to fitting the
line spectrum.  We adopted the median intrinsic width of the C$^{18}$O lines as the best estimate of the 
line width and fixed this parameter in the fits to individual lines. The measured mean FWHMs of the absorption
profiles are 4.5\,$\rm km\,s^{-1}$ for Reipurth 50 and 5.1\,$\rm km\,s^{-1}$ for VV CrA, which, when deconvolved with the instrumental profile,
yield $\gamma_{\rm intrinsic}$ = 3.3 and 4.0\,$\rm \ km\,s^{-1}$, respectively. The scatter of measured line widths
suggests that the uncertainty on the median width is 10-20\%. The value of $\gamma_{\rm intrinsic}$ is related
to the Doppler velocity dispersion, $b$, by  $\gamma_{\rm intrinsic} = 2\sqrt{\ln 2}\;b$. 
The baseline intensities, $I_c$, were fit simultaneously with the line profiles using first-order polynomial fits to the local continuum. Sample fits are illustrated in Figure \ref{specfits}. 
%

\begin{table*}
\centering
\caption{Measured CO isotopologue rotational temperatures.\tablenotemark{a}}
\tablenotetext{ a}{Values were obtained from a simultaneous two-temperature model. All temperatures in K.}
\begin{tabular}{lllll} 
\hline
\hline
& VV CrA - high T & VV CrA - low T & Reipurth 50 - high T & Reipurth 50 - low T \\
\hline

$^{12}$C$^{16}$O& $261\pm 5$ \,\, & $16\pm 1$ \,\,& $152\pm 9$ \,\,&$19\pm 3$ \,\, \\
$^{13}$C$^{16}$O& $>100$     \,\, & $16\pm 2$ \,\,& $212\pm 4$\,\,&$18\pm 1$ \,\, \\
$^{12}$C$^{18}$O& $176\pm 11$\,\, & $\ \ 6\pm 1$ \,\,& $150\pm 3$ \,\,&$21\pm 2$\,\, \\
$^{12}$C$^{17}$O& $173\pm 30$\,\,& $15\pm 5$ \,\,& $116\pm 6$ \,\,&$16\pm 3$ \,\, \\
\hline
\end{tabular}
\label{temp_table}
\end{table*}

\begin{table*}
\centering
\caption{Measured CO isotopologue column densities and ratios.  Uncertainties reflect the 68\% confidence level}
\begin{tabular}{llllll}
\hline
\hline
& VV CrA - high T & VV CrA - low T\tablenotemark{a} & Reipurth 50 - high T & Reipurth 50 - low T\tablenotemark{a}  & Local ISM\tablenotemark{b} \\
\hline
N($^{12}$C$^{16}$O)              & $7.2\pm0.2 \times 10^{18}\rm\,cm^{-2}$    & $1.4\pm 0.1 \times 10^{18}\rm\,cm^{-2}$& $9.0\pm 0.5\times 10^{18}\rm\,cm^{-2}$& $1.6\pm 0.2 \times 10^{18}\rm\,cm^{-2}$ &$ $  \\
N($^{12}$C$^{16}$O)/N($^{13}$C$^{16}$O)& $100\pm10\tablenotemark{c}$& $110\pm 8$   & $110\pm 7$    & $65 \pm 7$	&$69 \pm 6$   \\
N($^{12}$C$^{16}$O)/N($^{12}$C$^{18}$O)& $690\pm 30 $              & $1700\pm 250$ & $490\pm 30$   & $340 \pm 40$  &$557 \pm 30$	  \\
N($^{12}$C$^{16}$O)/N($^{12}$C$^{17}$O)& $2800\pm 300 $            & $4200\pm 800$& $2200\pm 150$ & $2300 \pm 340$	  &$2005 \pm \,155$  \\
N($^{12}$C$^{18}$O)/N($^{12}$C$^{17}$O)& $4.1\pm 0.4$              & $2.5\pm 0.6 $& $4.4\pm 0.2$  & $6.7\pm 0.8$  &$3.6 \pm\, 0.2$  \\
\hline
\end{tabular}
\tablenotetext{a}{Low-temperature values are shown for completeness; see text for discussion on their uncertainties.}
\tablenotetext{b}{\citep{Wilson99}}
\tablenotetext{c}{assuming $T(^{12}{\rm CO})=T({^{13}{\rm CO}}$)}
\label{ratio_table}
\end{table*}

From these fits we obtained rotational diagrams for the respective sources, shown in Figure \ref{rotplot}. These plots show the sub-level column densities, $N_{J}/(2J + 1)$, derived from the fits to the absorption lines, versus the energy of the state $J$. A linear trend is expected for a single-temperature gas. It is clear from the different slopes between the low-$J$ transitions ($J \le 3$) and higher-$J$ transitions that there is a range of different gas temperatures along the line of sight toward both sources. This is to be expected for essentially all geometries in which the gas is heated by the central star. While assuming a continuous distribution of temperatures might be most realistic, we found that a distribution consisting of only two temperatures is sufficient to capture the distribution of points in the excitation diagrams for both sources. In the absence of any explicit information on the actual temperature distributions, we proceeded with a simultaneous two-temperature fit to the data for each isotopologue (Figure \ref{rotplot}); from these models, individual low and high temperatures were derived. For both objects, the two temperatures consisted of a cold component of $T\sim 10-20\,$K and a warm component of $T\sim 150-250\,$K.   
In the case of VV CrA, it is worth noting that the temperature of the warm component is roughly consistent with that of the uppermost surface layers at a few 100 AU in
detailed disk heating models \citep{Kamp04,Jonkheid2004}. The column density of the warm component is significantly larger than expected from these models perpendicular to the midplane, but may be
explained if small dust grains have been depleted from the upper disk layers with a concommitant drop in opacity. A non-vertical, slanting line of sight through the disk can also enhance the total column of warm gas. Further modeling of these specific sources is needed to determine if the observed temperatures and column densities are quantitatively consistent with flared disk models. 
 
In both VV CrA and Reipurth 50, the derived low temperatures are consistent with the outer disk near the midplane or, in the case of Reipurth 50, possibly the surrounding envelope.  The high-temperature contributions to the measured absorption strengths of lines dominated by the cold component are considerable (Figure \ref{rotplot}). 
  
\begin{figure*}
\plottwo{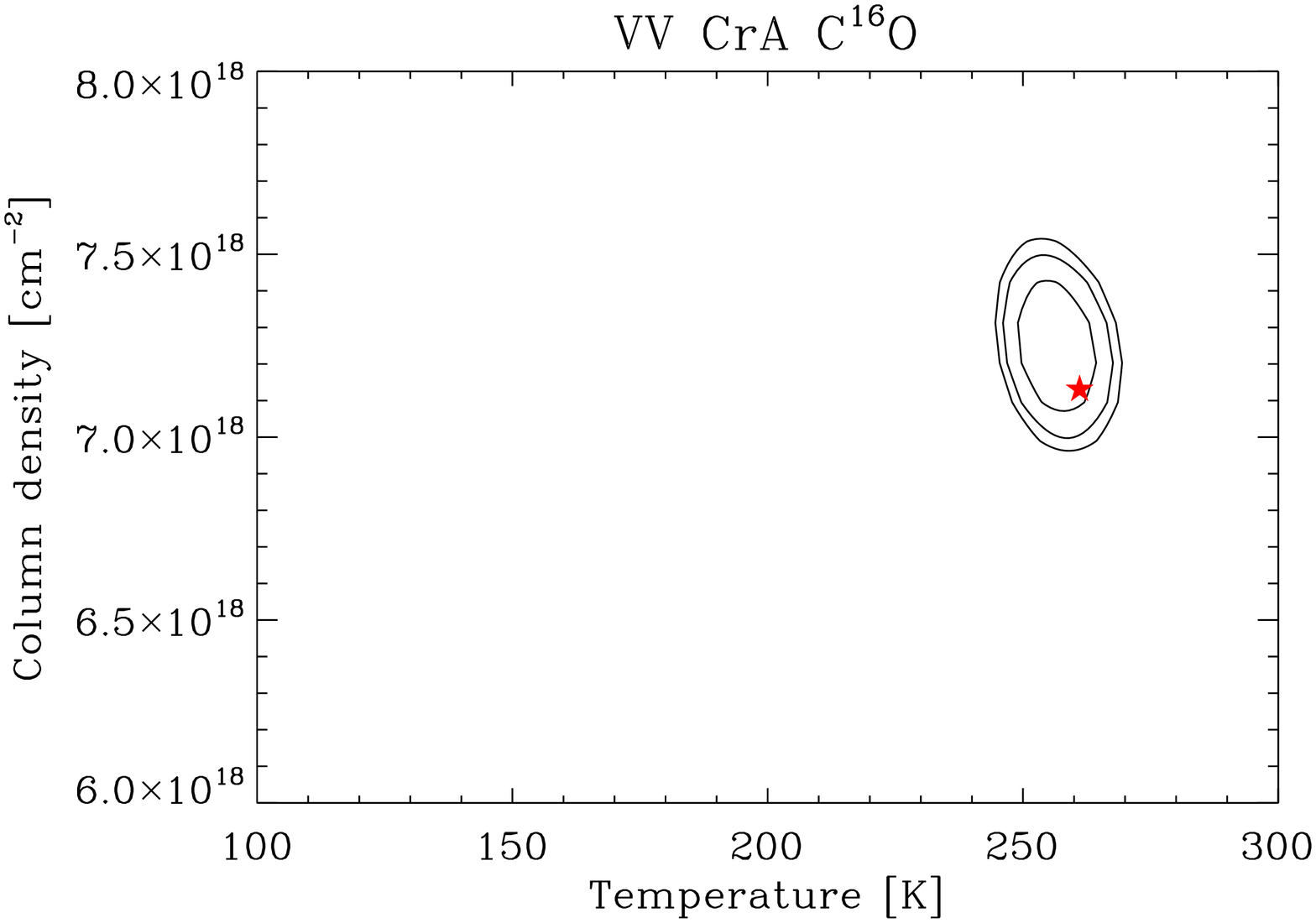}{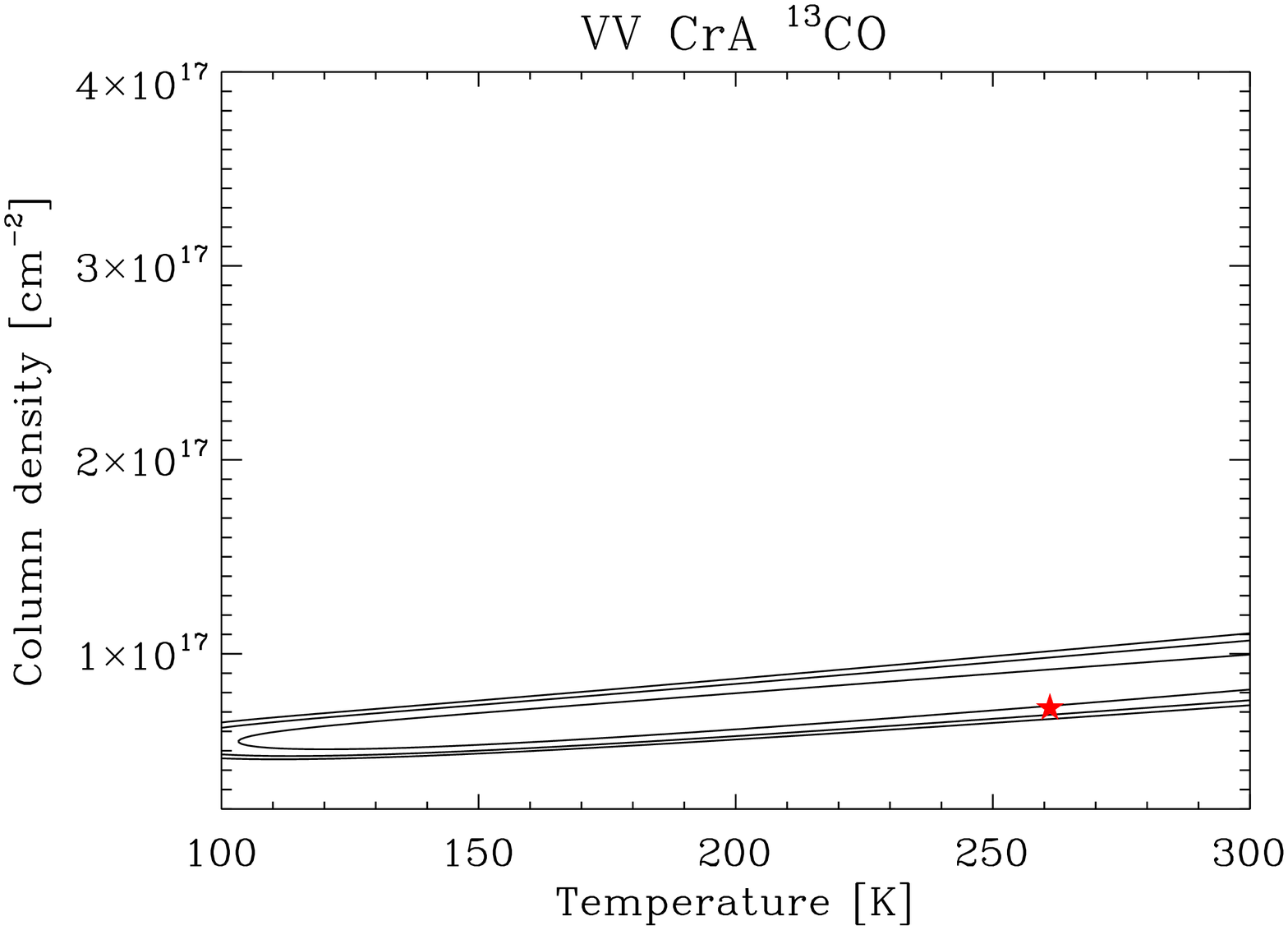}
\plottwo{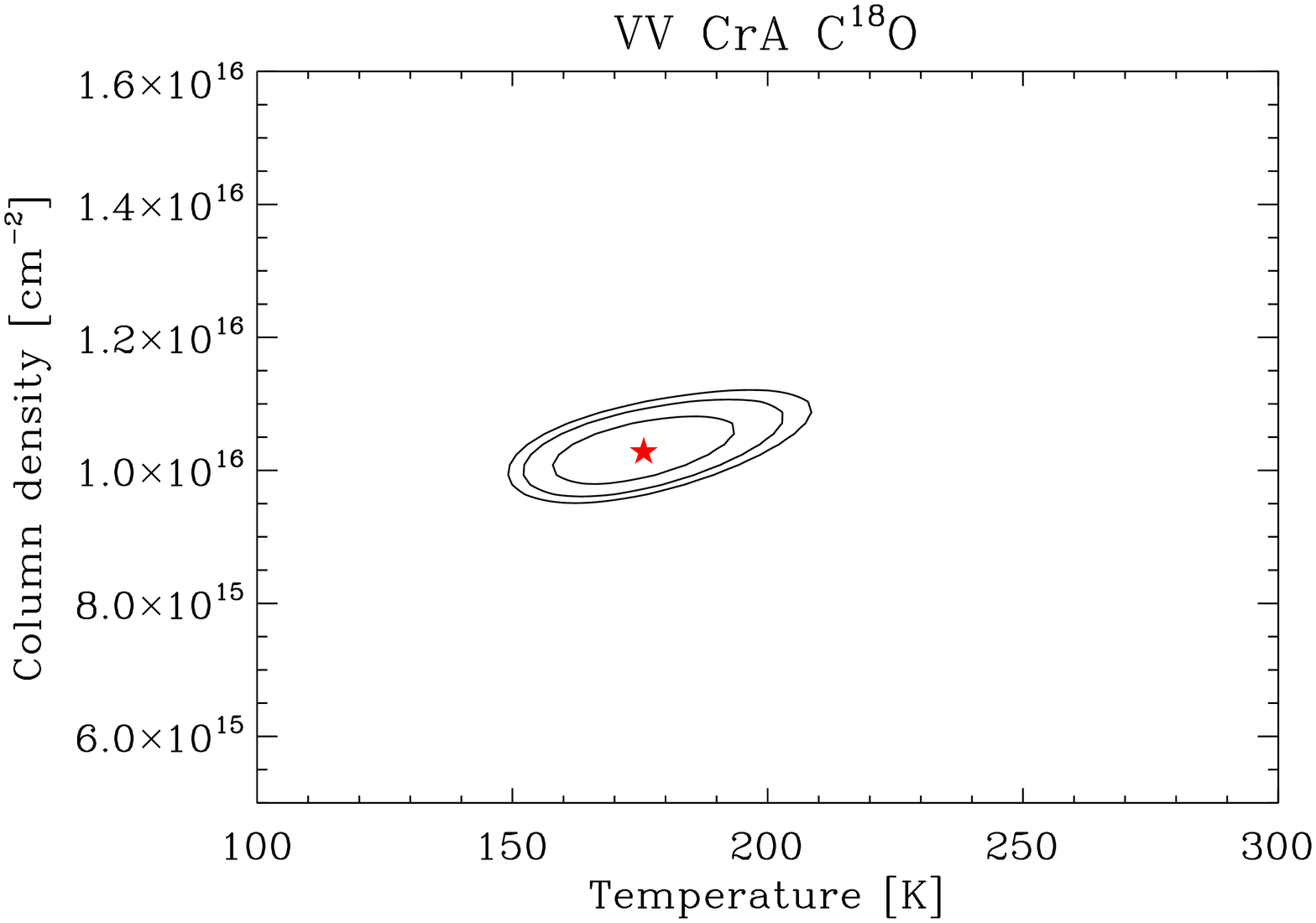}{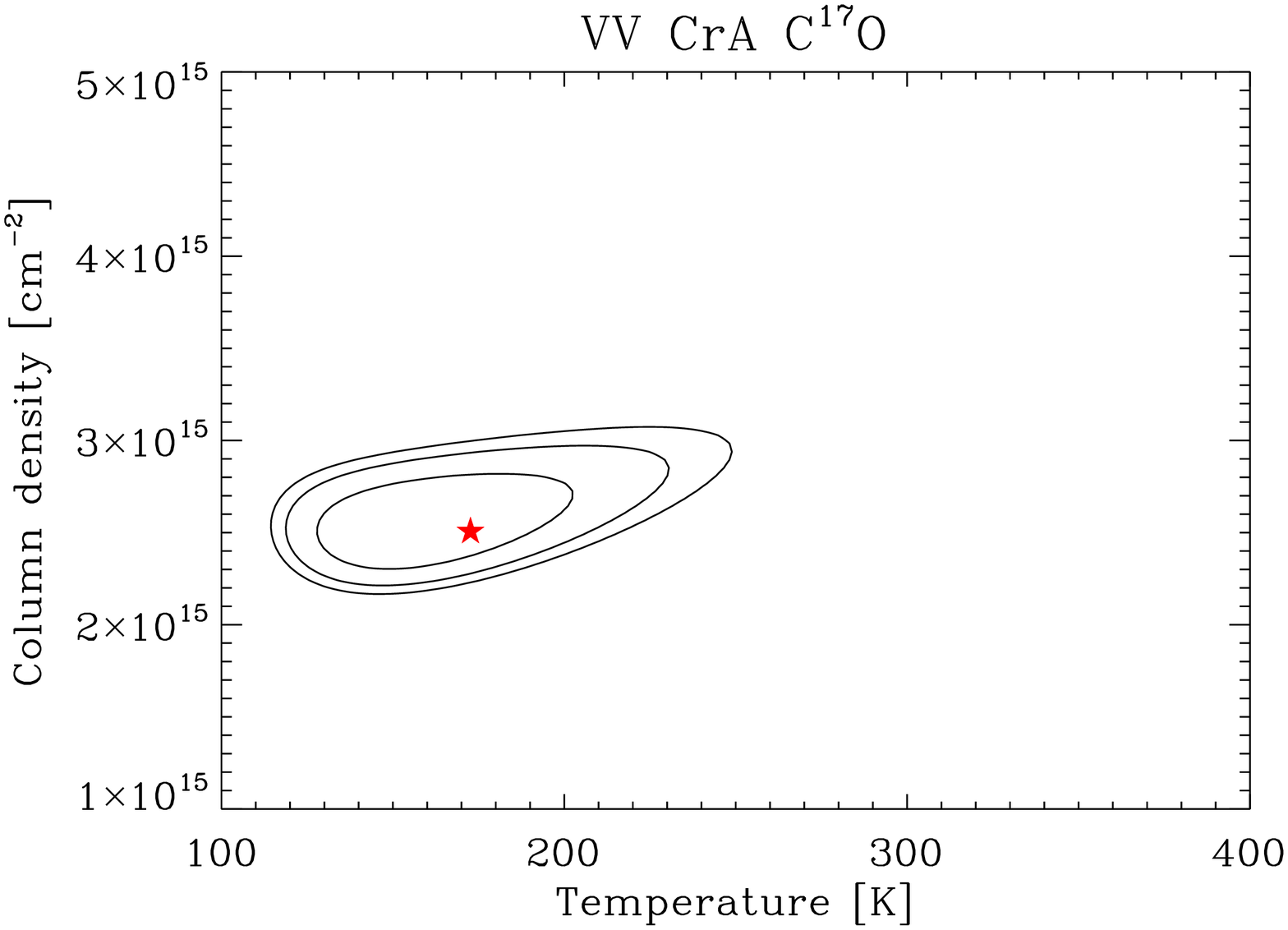}
\caption[]{Goodness-of-fit contours for the high temperature lines ($J\ge 4$) of the four measured isotopologues in VV CrA. 
The contours show the 68, 95, and 99\%  confidence levels. Red stars indicate total column densities obtained from the simultaneous two-temperature model. }
\label{goodnessfit_vvcra_hiT}
\end{figure*}

\begin{figure*}
\plottwo{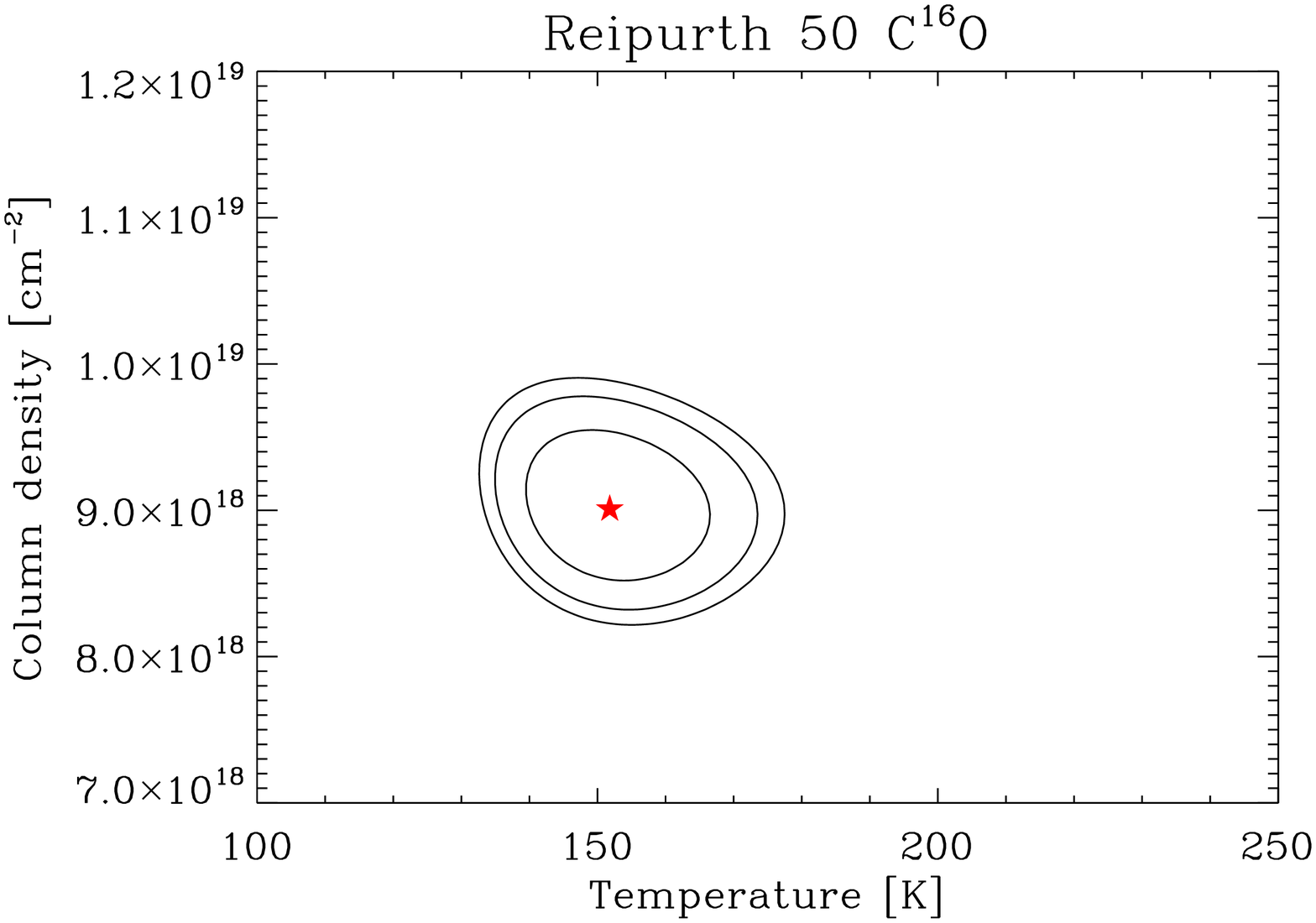}{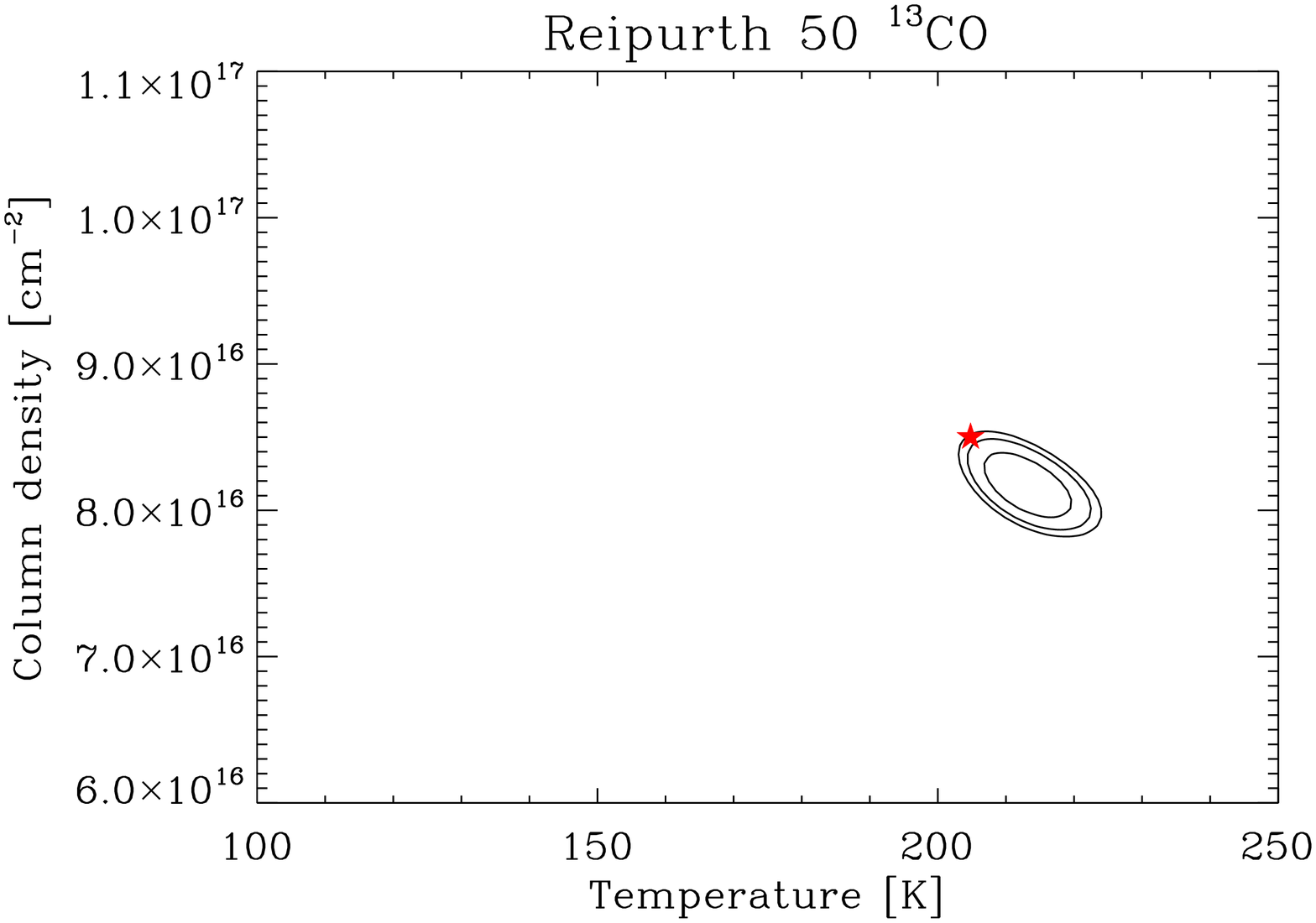}
\plottwo{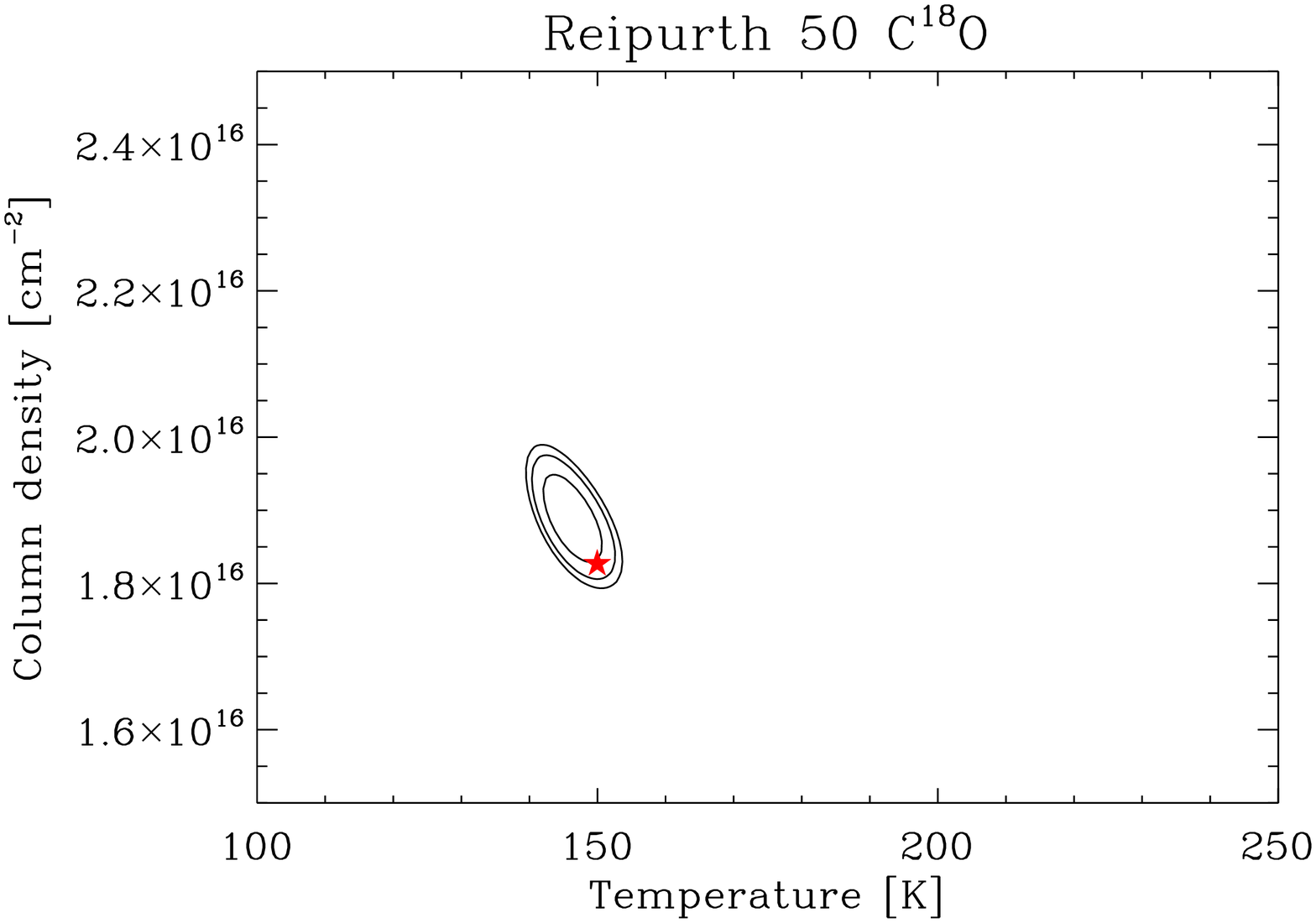}{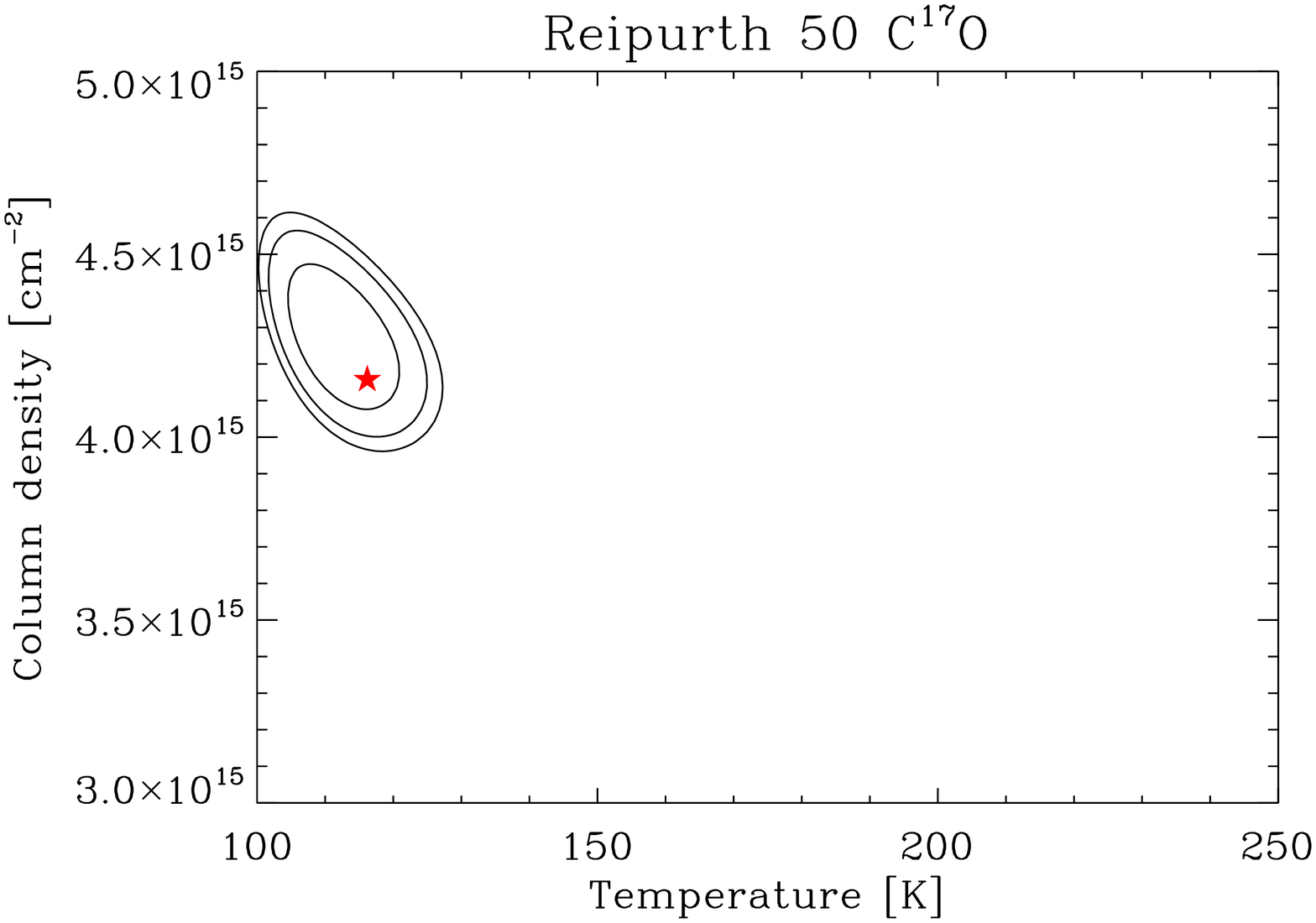}
\caption[]{Goodness-of-fit contours for the high-temperature lines ($J\ge 4$) of the four measured isotopologues in Reipurth 50. 
The contours show the 68, 95, and 99\%  confidence levels. Red stars indicate total column densities obtained from the simultaneous two-temperature model.}
\label{goodnessfit_re50_hiT}
\end{figure*}

The uncertainties in individual line depths are dominated by systematics, and these are taken into account
in the final errors for the isotopologue ratios. The greatest source of systematic uncertainty is likely to be residuals from telluric lines that were not completely removed. 
This source of error is difficult to quantify, and varies strongly for different lines. To account for this, we added 5\% uncertainties in quadrature to each line strength for most of the isotopologues in both temperature regimes; exceptions to the 5\% uncertainty were accounted for in VV CrA for the low temperature $^{12}$C$^{17}$O, low temperature $^{12}$C$^{18}$O and high temperature $^{13}$C$^{16}$O lines, for which no systematic uncertainty, 10 \% and 20 \% systematic uncertainty were included, respectively. These variations resulted from variable optimizations in fits to obtain reduced chi-square values near unity; chi-squared ellipses are shown in the temperature-column density plane (Figures \ref{goodnessfit_vvcra_hiT} and \ref{goodnessfit_re50_hiT}).
A few lines are clearly outliers relative to a linear fit in the rotational plot by many standard deviations; these lines, shown as faded points in the fits in Figure \ref{rotplot}, were 
excluded from the final fits. Inclusion of the estimates for systematic errors means the uncertainties cited here are likely to be conservative.

\subsection{Isotopic Ratios and Uncertainties}

Derived excitation temperatures and isotopologue ratios are listed in Tables \ref{temp_table} and \ref{ratio_table}, respectively.
There is a significant difference in excitation temperatures among the different isotopologues as evidenced by the variable slopes in Figure \ref{rotplot} for each temperature component, apparently violating the assumption of local thermodynamic equilibrium. For example, the excitation temperature for the high-temperature $^{13}$CO component in Reipurth 50 is greater than that of the other isotopologues (Figure \ref{goodnessfit_re50_hiT}). A similar effect is seen in the warm VV CrA $^{12}$CO lines (Figure \ref{goodnessfit_vvcra_hiT}). However, the apparently higher temperature of $^{12}$CO could be due
to line trapping in optically thick ro-vibrational lines, in which energy escapes from the rarer isotopologues more readily than from $^{12}$CO due to high optical depth in the latter. 
In this case, the derived isotopologue ratios should not be affected. 

Because the $^{12}$CO lines are observed in a different wavelength region (2.3\,$\mu$m) than the other isotopologues (4.7\,$\mu$m), it is possible that
the source location and size are not exactly the same due to the greater scattering of the K band relative to the M band. This could cause significant differences in the column densities that are probed. If the absorbing gas is very close to an infrared emitting source that changes in size or position with wavelength, it is still possible that we are probing different volumes and perhaps path lengths. However, in a spherical envelope, material with temperatures of 150-200 K is located at distances of $\sim$100-200\ AU from a 250\,$L_{\odot}$ source, compared
to a size of only a few AU for the infrared source. Hence, this is potentially the largest source of error 
for both an embedded source with a large reflection nebula like Reipurth 50, as well as a single inclined disk geometry as illustrated by Case B (Figure \ref{diskcartoon}) for VV CrA. The more probable two-disk scenario for VV CrA (Case A) is not likely to suffer from optical path differences.  With CRIRES, the pointing is the same for all spectral settings, with both K- and M-band observations tracking sources in the K band. We therefore believe that there are no significant differences in the absorbing volumes for the K and M bands for Reipurth 50 and VV CrA.

The low temperatures derived from our model are low enough that they could well correspond to the temperature of the surrounding molecular cloud for Reipurth 50, or the outer disk in Case A or B for VV CrA, perhaps weakly heated by the nearby protostar. However, due to the significant contribution of the high-temperature component on the low-temperature lines, we believe that our low-temperature ratios contain a much larger degree of uncertainty than we are able to precisely quantify. We therefore do not believe these numbers are meaningful. In discussing isotope ratios, we therefore refer to the ratios obtained from the high-temperature lines only. 
\

\section{Discussion}

Both VV CrA and Reipurth 50 should have $\rm [C^{16}O]/[C^{18}O]$ and $\rm [C^{16}O]/[C^{17}O]$ 
values similar to the local ISM in the absence of extensive isotope partitioning by photochemistry 
(self-shielding) or mass-dependent kinetics. The oxygen isotope ratios for the ISM as a function of distance from the Galactic center (Galactocentric radius, $R$$_{\rm GC}$) can be estimated from the  
4 kpc and 8 kpc values reported by \cite{Wilson99}, yielding for $\rm [^{16}O]/[^{18}O]$:

\begin{equation}
{\rm[ ^{16}O]/[^{18}O]}_{R_{\rm GC}} = (57.5 \pm 10)\  {R_{\rm GC}} + 97,
\label{eq:5}
\end{equation}

\noindent where the $1\sigma$ uncertainty in the slope reflects the uncertainty in the means (standard errors) reported in Table 4 of  \citep{Wilson99}. For $\rm [^{16}O]/[^{17}O]$ we can combine equation \ref{eq:5} 
with values for $\rm [^{18}O]/[^{17}O]$ in the ISM.  These estimates vary from 3.5 \citep{Wilson99} to 4.1 \citep{Wouterloot2008-487}. 

 Based on equation \ref{eq:5} and  $\rm [^{18}O]/[^{17}O] = 4.1$ for the ISM, one should expect CO surrounding VV CrA ($R_{\rm GC} = 7.8$\,kpc) to have $\rm [C^{16}O]/[C^{18}O]$ of $550 \pm 90$ 
and $\rm [C^{16}O]/[C^{17}O]$ of $2300 \pm 250$. The measured values of $690 \pm 30 (1\sigma)$ and $2800 \pm 300$ for VV CrA 
are greater.  Conversely, the expected $\rm [C^{16}O]/[C^{18}O]$ and $\rm [C^{16}O]/[C^{17}O]$ 
values for Reipurth 50 ($R_{\rm GC} = 8.4$\,kpc), $580 \pm 90$ ($1\sigma$) and $2400 \pm 270$, respectively, 
are indistinguishable from the measured values of $490 \pm 30$ ($1\sigma$) and $2200 \pm 150$ (Table \ref{ratio_table}). 

\begin{figure}
\includegraphics[width=8.5cm]{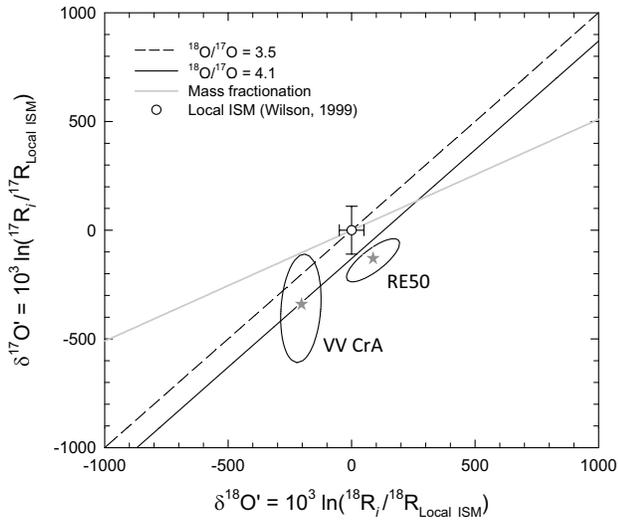}
\caption[]{Comparison of oxygen isotope ratios between the interstellar medium and CO surrounding the two YSOs
reported here (stars). Ratios for the YSOs are derived from the high-temperature CO abundances. The solid and dashed lines show mass-independent fractionation
lines with different assumptions for the $^{18}$O/$^{17}$O ratio, as indicated. The grey line shows the
mass-dependent fractionation line. Ellipses represent 95\% confidence limits derived from the goodness-of-fit contours shown in Figures \ref{goodnessfit_vvcra_hiT} and \ref{goodnessfit_re50_hiT}.}
\label{three_isotope_plot}
\end{figure}

Results for the oxygen isotopologues in the high-temperature regime are summarized in the three-isotope plot shown in Figure \ref{three_isotope_plot}, where differences in $\rm [^{16}O]/[^{18}O]$ 
and $\rm [^{16}O]/[^{17}O]$ from local ISM values are shown in per mil using the 
linearized form of the delta notation commonly used in isotope cosmochemistry. Here, $\delta^{18}{\rm O}^{\prime} = 10^3
\ln(^{18}R_i/^{18}R_{\rm Local\ ISM})$ 
and $^{18}R_i$ = $\rm [C^{18}O]/[C^{16}O]$. Error ellipses 
in the three-isotope plot were calculated using a Monte Carlo calculation that propagates uncertainties in column densities to 
the final delta values. The three-isotope plot  
shows that the differences in $\rm [^{16}O]/[^{18}O]$ and $\rm [^{16}O]/[^{17}O]$ between best estimates of the local ISM and the disk 
surrounding \object{VV CrA} are not consistent with mass-dependent isotope fractionation (represented by a line with a slope of 0.52 
in Figure \ref{three_isotope_plot}). This conclusion is firm to the extent that it can be assumed that the precursor molecular 
cloud to VV CrA was similar to the local ISM. The conclusion is robust despite the uncertainty in the $\rm [^{18}O]/[^{17}O]$ ratio 
of the ISM because the position of the \object{VV CrA} data point on the higher end of the range in reported $\rm [^{18}O]/[^{17}O]$ 
values (near 4.1) precludes mass-dependent fractionation as a primary control on the CO isotopologue ratios in all cases.   
Rather, the trajectories between VV CrA and plausible local ISM isotopic compositions have slopes of 1 or greater. 
The slope-1 relationship between local ISM and CO in the VV CrA disk is reminiscent of the mass-independent fractionation that 
characterizes oxygen in the solar system and is a tell-tale sign of self-shielding during CO photolysis.

The ability to distinguish 
slope-1 (mass-independent) from slope-0.52 (mass-dependent) trends in Figure \ref{three_isotope_plot} is critical for 
interpretation of the data and is a key advantage of having data for all three oxygen isotopologues of CO. The deficit of the rarer isotopologues, $\rm C^{18}O$ and $\rm C^{17}O$, relative to $\rm C^{16}O$ in the VV CrA disk,
is on the order of 20 to 40\% respectively, which is significant at the $>2\sigma$ level.   

The [$^{12}$CO]/[$^{13}$CO] ratios for both VV CrA and Reipurth 50 are similar, with high-temperature values of $100\pm 10$ and $110\pm 7$, respectively (Table \ref{ratio_table}). Ratios for both objects are high compared with
the local ISM value of $69\pm 6$ \citep{Wilson99}. In a case where CO self-shielding was
almost certainly responsible for
extreme C$^{16}$O excesses in the study of the X Persei molecular cloud by \cite{Sheffer2002-L171}, there are no associated
excesses in $^{12}$C, suggesting
that carbon and oxygen isotope effects during CO photolysis can be decoupled.  Because both of our objects
have the same $[^{12}$C]/$[^{13}$C] but one has an excess in $\rm C^{16}O$
relative to the ISM, our data also suggest a decoupling  of
C and O isotope ratios in CO in these objects.
Self-shielding by CO
during photolysis should in principle result in an excess of
$^{12}$CO relative to $^{13}$CO. However, atomic carbon liberated by CO
photolysis is photoionized largely to C$^+$, and C$^+$ reacts rapidly to exchange carbon
with CO by the reaction $^{13}$C$^+$ + $^{12}{\rm C}$O $\rightleftharpoons$ $^{12}$C$^+$ + $^{13}$CO
\citep{Warin1996-535, VanDishoeck1988, Langer84}, hence driving the $^{13}$C
back into CO and diminishing the signature
of selective photodissociation. In contrast, oxygen
liberated by CO photolysis remains neutral in the photodissociation regions (PDRs)
and is relatively unreactive; oxygen will not exchange
with CO or other molecules prior to sequestration as H$_2$O
produced either in the gas phase by a reaction channel involving
H$_3^+$ \citep{Herbst2000-2523} or on grain surfaces \citep{Hasegawa1992-167}. 
The decoupling between carbon and oxygen effects seen here can be explained in the context
of photochemical self-shielding but offers as yet no explanation for the anomalously high [$^{12}$CO]/[$^{13}$CO] for both VV CrA and Reipurth 50. This anomaly will be the focus of a later publication.

The cause of the disparities in $\rm [^{16}O]/[^{18}O]$ and $\rm [^{16}O]/[^{17}O]$ between the disk 
 surrounding \object{VV CrA} 
and the local ISM cannot yet be deduced unequivocally, as more data points for Figure \ref{three_isotope_plot} are needed. However, the mass-independent
character of the displacement of this source is most consistent with a 
photochemical enhancement of $\rm C^{16}O$ relative to both $\rm C^{18}O$ and $\rm C^{17}O$
due to self-shielding by the more abundant $\rm C^{16}O$ isotopologue. 

The absence of a similar effect in Reipurth 50 might be attributable to the different geometry of the absorbing component of this source. The protostellar envelope in 
Reipurth 50 is presumably less affected by CO self-shielding than the disk in VV CrA. If the difference in mass-independent isotope fractionation between
disks and protostellar envelopes indeed holds for more objects, an
important question is what is the cause. The time-scale for the
isotope-selective photodissociation and subsequent chemistry itself should be
short, less than 1000 years, in the PDR of both types of objects. However, the lines
of sight likely probe more than just the PDR layer so the isotope
signature of these other layers is equally relevant. In the disks, the
photodissociation is confined to thin surface regions but the $\rm ^{16}O$-depleted
atomic oxygen can be mixed vertically down to the
midplane region on timescales of $\sim$10$^{5}$ years in the outer disk, where it
can then be sequestered into H$_{2}$O (Young 2007). Together with the
effects of a non-vertical geometry (\S 4.1) and grain growth by coagulation, this can
result in an isotopic fractionation effect over a much larger column
than would be expected from a thin PDR layer; grain growth can lead to higher UV flux through the disk and greater photodissociation.  In contrast, Reipurth 50, being only in stage I of its evolution, may not have had sufficient time to acquire a measurable $\rm C^{16}O$ excess due to mixing $\rm(\sim 10^{5} \ yr)$, so that, in the absence of mixing, 
the isotope fractionation of the PDR layer contributes little to the
total column.

\section{Conclusions}
Our new high-resolution IR absorption data for two YSO environments, a protoplanetary disk and a protostellar envelope, 
demonstrate that measurement of both $\rm[C^{16}O]/[C^{18}O]$ and $\rm [C^{16}O]/[C^{17}O]$ 
in CO is possible with sufficient precision to distinguish photochemical effects from mass-dependent isotope fractionation.  
The new high-resolution data exhibit a detectable deficit
in $\rm C^{18}O$ and $\rm C^{17}O$ in the disk surrounding VV CrA relative to the local interstellar medium. 
The CO surrounding Reipurth 50 exhibits no discernible differences in $\rm [C^{16}O]/[C^{18}O]$ 
and $\rm [C^{16}O]/[C^{17}O]$  relative to the local ISM.  One likely explanation for the different oxygen isotopologue
ratios for these objects is that a photochemical deficit in $\rm C^{17}O$ and $\rm C^{18}O$ relative to $\rm C^{16}O$  
proceeds only in a disk geometry and may require $\gtrsim 10^5$ years in the environments surrounding YSOs. Future analyses of more objects in different stages of evolution are now needed to enable differentiation between potential self-shielding in the disk versus the parent cloud.

\acknowledgments{
We thank the anonymous reviewer for constructive and valuable comments that improved this paper. Support for KMP was provided by NASA through Hubble Fellowship grant \#01201.01 
awarded by the Space Telescope Science Institute, which is operated by the Association of 
Universities for Research in Astronomy, Inc., for NASA, under contract NAS 5-26555. Astrochemistry in Leiden is supported by a Spinoza grant of the Netherlands Organization for
Scientific Research (NWO). This work was supported in part by a grant from the NASA Origins program (EDY, MM) and a grant from the NASA Astrobiology Institute (UCLA lead team). RLS was supported by these NASA Origins and Astrobiology grants. 
}

\bibliographystyle{apj}
\bibliography{Smith_et_al}

\end{document}